\documentclass{emulateapj}
\usepackage{apjfonts}
\usepackage{epsf}
\begin{document}
\slugcomment{}
\shortauthors{J. M. Miller et al.}
\shorttitle{NGC 4151}

\title{X-ray Structure between the innermost disk and optical broad line
  region in NGC 4151}

\author{J.~M.~Miller\altaffilmark{1},
E.~Cackett\altaffilmark{2},
A.~Zoghbi\altaffilmark{1},
D.~Barret\altaffilmark{3,4},
E.~Behar\altaffilmark{5},
L.~W.~Brenneman\altaffilmark{6},
A.~C.~Fabian\altaffilmark{7},
J.~S.~Kaastra\altaffilmark{8,9,10},
A.~Lohfink\altaffilmark{11},
R.~F.~Mushotzky\altaffilmark{12},
K.~Nandra\altaffilmark{13},
J.~Raymond\altaffilmark{6}
}

\altaffiltext{1}{Department of Astronomy, University of Michigan, 1085
  South University Avenue, Ann Arbor, MI 48109-1107, USA,
  jonmm@umich.edu}
\altaffiltext{2}{Department of Physics \& Astronomy, Wayne State
  University, 666 West Hancock Street, Detroit, MI, 48201, USA}
\altaffiltext{3}{IRAP CNRS, 9 Av. colonel Roche, BP 44346, F-31028
  Toulouse cedex 4, France}
\altaffiltext{4}{Universite de Toulouse III Paul Sabatier / OMP,
  Toulouse, France}
\altaffiltext{5}{Department of Physics, Technion, 32000,
  Haifa, Israel}
\altaffiltext{6}{Harvard-Smithsonian Center for Astrophysics, 60
  Garden Street, Cambridge, MA, 02138, USA}
\altaffiltext{7}{Institute of Astronomy, University of Cambridge, CB3
  OHA, UK}
\altaffiltext{8}{SRON Netherlands Institute for Space Research,
  Sorbonnelaan 2, 3584 CA Utrecht, Netherlands}
\altaffiltext{9}{Leiden Observatory, Leiden University, PO Box 2300 RA
  Leiden, Netherlands}
\altaffiltext{10}{Department of Physics and Astronomy, Universiteit
  Utrecht, PO BOX 80000, 3508 TA Utrecht, Netherlands}
\altaffiltext{11}{Department of Physics, Montana State University,
  Barnard Hall, Bozeman, MT 59717, USA}
\altaffiltext{12}{Department of Astronomy, University of Maryland
  College Park, MD, 20742, USA}
\altaffiltext{13}{Max Planck Intitute for Extraterrestrial Physics,
  Giessenbachstrasse, D-85741, Garching, Germany}

\begin{abstract}
We present an analysis of the narrow Fe~K$\alpha$ line in {\em
  Chandra}/HETGS observations of the Seyfert AGN, NGC 4151.  The
sensitivity and resolution afforded by the gratings reveal {\it
  asymmetry} in this line.  Models including weak Doppler
boosting, gravitational red-shifts, and scattering are generally
preferred over Gaussians at the $5\sigma$ level of confidence, and
generally measure radii consistent with $R\simeq
500-1000~GM_{BH}/c^{2}$.  Separate fits to ``high/unobscured'' and
``low/obscured'' phases reveal that the line originates at smaller
radii in high flux states; model-independent tests indicate that this
effect is significant at the 4--5$\sigma$ level.  Some models
and $\Delta t\simeq 2\times 10^{4}$~s variations in line flux suggest
that the narrow Fe~K$\alpha$ line may originate at radii as small as
$R\simeq 50-130~GM_{BH}/c^{2}$ in high flux states.  These results
indicate that the {\em narrow} Fe~K$\alpha$ line in NGC 4151 is
primarily excited in the innermost part of the optical broad line
region (BLR), or X-ray BLR.  Alternatively, a warp could provide the
solid angle needed to enhance Fe~K$\alpha$ line emission from
intermediate radii, and might resolve an apparent
discrepancy in the inclination of the innermost and outer disk in NGC
4151.  Both warps and the BLR may originate through radiation
pressure, so these explanations may be linked.  We discuss our results in
detail, and consider the potential for future observations with {\em
  Chandra}, {\em XARM}, and {\em ATHENA} to measure black hole masses
and to study the intermediate disk in AGN using narrow
Fe~K$\alpha$ emission lines.
\end{abstract}

\section{Introduction}
In many respects, NGC 4151 is a normal, if not a standard or archetype
Seyfert-1 active galactic nucleus (AGN).  The host galaxy is a
(weakly) barred Sab at a redshift $z=0.00332$ (Bentz et al.\ 2006), or
approximately 19 Mpc via dust parallax (Honig et al.\ 2014).  The
spectra and variability of the radiation produced from accretion onto
the central black hole have been studied extensively.  Indeed, NGC
4151 was among the first Seyfert-I AGN wherein strong correlations
between continuum bands and high ionization lines were detected (e.g.,
Gaskell \& Sparke 1986).  The 5 light-day delay detected in early data
is echoed in recent efforts: Bentz et al.\ (2006) measure a delay of
$6.6^{+1.1}_{-0.8}$~days between the 5100\AA~ continuum and H$\beta$
line, giving a ``broad line region'' (BLR) reverberation mass for the
black hole of $M_{BH} = 4.6^{+0.6}_{-0.5}\times 10^{7}~M_{\odot}$.  An
updated estimate, including the latest information on scale factors in
reverberations studies, gives $M_{BH} = 3.6^{+0.4}_{-0.4}\times
  10^{7}~M_{\odot}$ (Bentz \& Katz 2015).

NGC 4151 also appears to be a normal Seyfert-I in terms of its radio
properties.  It is ``radio-quiet'', meaning that any jet emission is
relatively unimportant compared to the energetic output of the
accretion disk.  A combination of VLBA and large-dish radio campaigns
have studied the central region of NGC 4151 at high angular resulotion
and high sensitivity.  The central source has a flux density of just
3~mJy, and a size less than 0.035~pc (Ulvestad et al.\ 2005); this is
only 10 times the size of the optical ``broad line region'' and much
smaller than a typical pc-scale torus geometry.  On arcsecond scales,
the radio emission takes the form of a two-sided radio jet, but the
outflow speed is constrained to be $v_{jet} \leq 0.16c$ and $v_{jet}
\leq 0.028c$ for distances of 0.16 and 6.8~pc from the nucleus
(Ulvestad et a.\ 2005).

NGC 4151 is also understood well on larger scales.  {\it Hubble}
spectroscopy strongly suggests that the optical narrow line region
(NLR) takes the form of a biconical outflow, at an inclination of
$45^{\circ}\pm5^{\circ}$ relative to our line of sight (Das et
al.\ 2005).  This inclination is consistent with the fact that both
sides of a low-velocity radio jet are observed; higher inclinations
and/or higher velocities might boost the counter jet out of the
observed band.  The NLR is also resolved in soft X-ray images obtained
with {\it Chandra}.  The diffuse X-ray emission appears to be
dominated by photoionization from the nucleus, and the overall
morphology and energetics are consistent with a biconical outflow with
a kinetic luminsosity that is just 0.3\% of the bolometric radiative
luminosity of the central engine (e.g., Wang et al.\ 2011).  

X-ray emission from the central engine in NGC 4151 paints a different
picture than that obtained in other wavelengths, and on larger scales.
The soft X-ray spectrum of NGC 4151 is sometimes heavily obscured;
complex, multi-zone absorption is required in these phases (e.g.,
Couto et al.\ 2016, Beuchert et al.\ 2017).  Spectra such as this are
typically obtained from Seyfert-2 AGN, wherein the ``torus'' blocks
most of the direct emission from the central engine.  Since the
fraction of Compton-thin AGN in the local universe is about 0.5 and
the average viewing angle of a disk in three dimensions is $\theta =
60^{\circ}$ (e.g., Georgakakis et al.\ 2017), obscuration is likely
tied to higher viewing angles.  The detection of highly obscured X-ray
spectra in NGC 4151 nominally indicates that its accretion flow is
observed at a {\it higher} inclination than indicated by its NLR.

It is unclear if simple inferences about viewing angles hold when the
obscuration is observed to be highly variable.  In NGC 4151, there are
phases wherein the continuum flux is higher and the obscuration is
greatly reduced.  If the torus is clumpy or irregular, and if our line
of sight to the central engine were to merely graze this structure,
periods of different obscuration might result.  The fact that changes
in NGC 4151 can occur fairly quickly -- within months and years --
indicates that at least some of the obscuration in NGC 4151 is closer
than torus.  Rather, it may originate as close as the optical BLR, or
even closer.  This may not be unique to NGC 4151; even in some bona
fide Seyfert-2 AGN, there are indications that the obscuration is much
closer than the torus (e.g. Elvis et al.\ 2004).

It has been apparent for some time that the optical BLR can be divided
into low-ionization and high-ionization regions (e.g., Collin-Souffrin
et al.\ 1988, Kolatchny et al.\ 2003).  Whereas the high-ionization
lines are often shifted with respect to the host frame and likely
originate in a wind, the low-ionization lines are typically not
shifted and may originate closer to the disk surface.  In at least
some cases (perhaps those viewed at low inclination angles),
double-peaked emission lines consistent with photoionization of the
accretion disk are observed, strongly suggesting a close association
with the disk itself (e.g., Eracleous \& Halpern 2003).  This
complexity may be echoed in X-ray studies.  Some highly ionized X-ray
``warm absorbers'' appear to be coincident with the optical BLR (e.g.,
Mehdipour et al.\ 2017).  However, X-rays also indicate that cold,
dense gas is cospatial in this region.  A systematic study of narrow
Fe K$\alpha$ lines observed with the {\it Chandra}/HETGS was
undertaken by Shu et al.\ (2010); the width of the observed lines
varies considerably, but most lines signal contributions from radii
consistent with the BLR.

Studies that probe the regions closest to the black hole in NGC 4151
-- near to the innermost stable circular orbit (ISCO) -- give a
different view of this AGN.  Recent efforts to study this regime using
time-averaged, flux- and time-selected intervals, and even
reverberation lag spectra demand that the innermost disk is viewed at
a {\em low} inclination with respect to the line of sight.  The most
complete analysis, including data from several missions and models
that allow for complex, multi-zone absorption at low energy, measures
an inclination of just $\theta = 3^{+6}_{-3}$~degrees (Beuchert et
al.\ 2017; also see Keck et al.\ 2015, Cackett et al.\ 2014, and
Zoghbi et al.\ 2012).

The simplest picture that can be constructed from AGN unification
schemes (e.g., Antonucci 1993, Urry \& Padovani 1995, Urry 2003) is
likely one where the angular momentum vectors and/or symmetry axes of
the black hole, accretion disk, BLR, torus, and NLR are all aligned.
Potential causes of misalignment include super-Eddington accretion
(e.g., Maloney, Begelman, \& Pringle 1996) and binarity (e.g, Nguyen
\& Bogdanovic 2016).  Therefore, evidence of misalignment in a given
AGN is potentially an important clue to recent changes in the
accretion flow and/or the recent evolution of the AGN.  If a
corresponding warp or transition region exposes more cold gas to the
central engine than a flat disk would at the same radius, a
characteristic Fe~K emission line may be produced.  The same warp may
help to launch a wind, providing a potential connection to the BLR.

Motivated by the need to better understand the nature and origin of
the BLR in Seyferts, and by the potentially linked question of a
misalignment of the innermost and outer accretion flows in NGC 4151,
we have undertaken a dedicated study of the narrow Fe~K$\alpha$ line
in this source.  In order to achieve the best possible combination of
sensitivity and resolution, we have restricted our analysis to
obserations made using the {\it Chandra}/HETGS.  In section 2, we
detail the observations that we have considered, and the methods by
which the data are reduced.  Section 3 summarizes the different
analyses that are attempted using a variety of techniques, models, and
data selections, and the results.  In Section 4, we discuss these
results in a broader context and highlight potential future
investigations.

\section{Observations and Reduction}
All of the archival {\it Chandra}/HETGS observations of NGC 4151 were
downloaded directly from the {\it Chandra} archive.  The observation
identification number (ObsID), start date, and duration of each
exposure are listed in Table 1.  HETGS observations nominally deliver
spectra from both the high energy gratings (HEG) and medium-energy
gratings (MEG).  The MEG has less effective area in the Fe K band and
lower resolution, and is therefore less suited to our analysis.
Moreover, flux uncertainties remain between the MEG and HEG. For
simplicity and self-consistency, then, we have limited our analysis to
the HEG.

Observations obtained directly from the archive include calibrated
``evt2'' files, from which spectra can be derived.  Using CIAO version
4.9 and the associated calibration files, we ran the following tools
on each ``evt2'' file to produce spectra: ``tgdetect'',
``tg\_create\_mask'', ``tg\_resolve\_events'', and ``tgextract''.
Within ``tg\_create\_mask'', we set the ``width\_factor\_hetg''
parameter to a value of 18 (half of its default value) in order to
better separate the first-order HEG and MEG spectra.  Instrument
response functions were generated using ``mkgrmf'' and ``fullgarf''
using the default parameters.

For each exposure, we added the first-order ($+1$ and $-1$) HEG
spectra and ancillary response files (ARFs) using the CIAO tool
``add\_grating\_spectra''.  Combined first-order HEG spectra from
different observations were added in the same manner.  The
redistribution matrix files (RMFs) for HEG spectra do not differ
between plus and minus orders, nor do they differ appreciably between
observations (the dispersive spectrometer has a diagonal response).
For all later spectral analysis, then, we arbitrarily chose an RMF
from one of the observations under consideration.

Spectra from different observations and intervals were created to suit
different needs, and binned according to their relative sensitivity.
Binning was accomplished using the tool ``grppha'' within the standard
HEASFOFT suite, version 6.19.  

All spectral fits were performed using XSPEC version 12.9 (Arnaud
1996).  When starting a given fit, the default ``standard'' weighting
was used.  This method allows the spectral model to quickly find a
good (but, merely good) fit to the data.  We then refined the fit
after adopting ``model'' weighting, which assigns an error to each bin
based on the counts predicted by the model.  This is slightly
more accurate than other weighting methods; in particular, it avoids
over-estimates of minus-side (toward zero flux) errors that are
intrinsic to ``standard'' weighting.  Unless otherwise noted, the
errors quoted in this work reflect the value of the parameter at its
$1\sigma$ confidence interval.  Errors were derived using the standard
XSPEC ``error'' command, and verified using the ``steppar'' command;
like ``error'', ``steppar'' varies all parameters jointly, but allows
the user to control how the parameter space is searched.

Light curves of individual observations were created using ``dmcopy'',
selecting events assigned to the first-order HETG extraction regions.
This procedure is preferable to light curves of the zeroth order,
owing to the potential for pile-up in the zeroth-order source image to
artificially mute variability.

\section{Analysis and Results}
\subsection{Spectral Fitting Range and Set-up}
Unless otherwise noted, the spectral fits presented in this paper are
restricted to the 6.0--6.7~keV range, in the observed frame.  This
band is sufficiently narrow that the local continuum can be
represented without complex models.  The lower bound of 6~keV is too
high to suffer substantial curvature due to low energy absorption, and
relativistic Fe~K lines from the inner disk stretch down to 4--5 keV
(e.g., Keck et al.\ 2015, Beuchert et al.\ 2017).  The sensitivity and
band pass of the HEG are not ideally suited to determining the
relative influence of multiple reflection and complex absorption
components that can broadly shape the spectrum above the neutral Fe~K
edge at 7.1~keV; the upper bound of our fitting range purposely avoids
this range.  Any Fe XXV emission or absorption lines near to 6.70~keV
are much weaker than the narrow Fe~K$\alpha$ line at 6.40~keV, and do
not affect the fit.

The point of adopting this narrow fitting range is that the modeling
results are driven by the {\it line}, leveraging the advantages of the
high spectral resolution currently afforded only by the HEG.
Functionally, this is similar to the manner in which optical and UV
spectra of AGN are analyzed: detailed fits are made to individual
lines with local, fidicial, continua.  In the limit of high resolution
and line sensitivity, no information is lost - dynamical and
scattering effects can be seen in a single line or a small collection
of lines.  This approach is possible in NGC 4151 since the Fe
K$\alpha$ line observed in this source is the brightest of any
Seyfert; in the calorimeter era, this approach will likely be a
pragmatic approach to the spectra of many sources.

The power-law parameters that are measured in our fits, then, do not
necessarily represent the true continuum emitted by the central
engine.  Rather, this spectral form is just a simple parameterization
of the local continuum, which may consist of direct emission,
reflection from the innermost disk, and distant reflection.  However,
we have still enforced some minimal restrictions on the power-law
index to ensure that it corresponds well to plausible values in AGN.
An index flatter than $\Gamma=1.4$ is unlikely to arise through
Comptonization (e.g., Haardt \& Maraschi 1993), so we have set this as
a lower bound.  Most Seyfert AGN appear to have power-law index of
$\Gamma=1.7-1.8$ (see, e.g., Nandra et al.\ 2007), and cases steeper
than $\Gamma=2.0$ appear to be restricted to narrow-line Seyfert-1s
and/or sources at or above the Eddington limit.  We have therefore set
$\Gamma=2.0$ as an upper bound on the power law index.

\subsection{Summed and State-specific Fits}
We initially considered fits to the summed spectrum from all
observations of NGC 4151 (see Table 1), and then constructed spectra
from periods with high flux and/or less obscuraton, and lower flux
and/or higher obscuration (hereafter and within Tables 2 and 3,
``high/unobscured'' and ``low/obscured'' states).  These distinctive
states of NGC 4151 are well known and they have been the subject of
various X-ray and multiwavelength studies (for a recent treatment, see
Couto et al.\ 2016).  Figure 1 shows the combined first-order HEG
spectra from each observation of NGC 4151 on the 1--10~keV band,
illustrating that the observations are easily grouped in this manner.
Observations with a count rate below 0.01~${\rm counts}~{\rm s}^{-1}$
were grouped into the ``low/obscured'' state, and those with a count
rate above 0.02~${\rm counts}~{\rm s}^{-1}$ were grouped into the
``high/unobscured'' state (see Table 1).

Prior to spectral fitting, the summed spectrum from all observations
was grouped to require 20 counts per bin.  The sensitivity in the
combined ``high/unobscured'' state spectrum is very similar, so it was
also grouped to require 20 counts per bin.  Depending on the specific
band and model used, the combined ``low/obscured'' state spectrum has
a continuum flux that is about 2--2.5 times lower than that in the
``high/obscured'' state.  Since flux errors scale as the square root
of the flux in each bin, a factor of $\sim5$ per bin must be recovered
to make a consistent comparison.  The ``low/obscured'' state spectrum
was therefore grouped to require 100 counts per bin prior to fitting.

The models that were applied to the spectra fall into two broad
categories: those that treat the Fe~K$\alpha$ line and continuum using
separate components (see Table 2), and reflection models that treat
the line and continuum jointly (see Table 3).  In all instances where
the statistical significance of an improved fit is quoted in the text
below, the signicance is based on the F-statistic probability for a
given difference in the fit statistic and number of free parameters
($\chi^{2}$ and $\nu$, respectively).  

\subsection{Gaussian Modeling}

The simplest models that we examined describe the Fe K$\alpha$ line
with a simple Gaussian function ($zgauss+zpow$, in XSPEC, where the
``$z$'' denotes that the components allow the redshift of the source to be
specified).  In all cases, we fixed the centroid energy of the
Gaussian at 6.40~keV in the source frame.  In our initial fits, the
line width was fixed at $\sigma = 0$~eV, effectively assuming a line
width below the instrumental resolution.  The next fits allowed the
width of the Gaussian to vary freely.  In the summed,
``high/unobscured'', and ``low/unobscured'' phases, the width is
consistent with $\sigma = 23$~eV, corresponding to a {\it projected}
velocity of $v\simeq 1100$~km/s.

Figure 2 shows the results of fitting Gaussian models to the summed,
``high/unobscured'', and ``low/obscured'' spectra.  The line is
clearly broader than the Gaussian model that only includes
instrumental broadening, signaling that the line is resolved.  The
narrow Gaussian fits illlustrate the asymmetry of the line, which has
a clear red wing.  The broadened Gaussian functions provide improved
fits, but these models are still not statistically acceptable (see
Table 2).  Particularly in the summed and ``high/unobscured'' spectra,
the broad symmetric functions effectively over-predict the blue wing of
the Fe~K$\alpha$ line while still failing to fully fit its red wing.

Here, it is important to note the neutral Fe~K$\alpha$ line is
actually composed of two lines, with lab energies of 6.391~keV and
6.404~keV (e.g., Bambynek et al.\ 1972).  The $\sigma = 23$~eV width
of the best-fit Gaussian models corresponds to a FWHM of 54~eV; the
separation of these lines is only 24\% of the measured FWHM of the
line, and therefore does not significantly skew our results.  The
separation of these lines is also below the nominal resolution of the
first-order HEG in the Fe K band (approximately 45~eV).  We have fit
only a single Gaussian for simplicity, and because this follows prior
work against which our results must be compared (e.g., Shu et
al.\ 2010; see below).  The next models that are considered represent
specific improvements over simple Gaussian functions; some allow for
much broader and asymmetric line profiles (e.g., ``diskline'', Fabian
et al.\ 1989 ), and others explicitly include both Fe~K$\alpha$ lines
within the model (e.g., ``mytorus'', Murphy \& Yaqoob 2009, Yaqoob \&
Murphy 2010).

\subsection{Diskline Modeling}

The line asymmetry indicated in Figure 2 suggests that dynamical
effects -- at some distance from the ISCO -- may shape the line
profile.  The next set of fits we made therefore replaced the Gaussian
model for the line with the ``diskline'' function (Fabian et
al.\ 1989).  This model describes the distortions to an intrinsically
symmetric Gaussian line emitted from the accretion disk around a
Schwarzschild black hole.  If the data contained extremely broad
lines, consistent with orbits close to the black hole, ``diskline''
would be inadequate because it assumes a spin of $a = 0$.  However,
newer line models that make black hole spin a free parameter, e.g.,
``kerrdisk'' and ``relline'' (Brenneman \& Reynolds 2006, Dauser \&
Garcia 2013) are only valid up to $400~GM/c^{2}$ and $1000~GM/c^{2}$,
respectively.  Far from the black hole, the difference between
Schwarzshild and Kerr metrics is unimportant; moreover, ``diskline''
is {\em analytic} and can easily be extended to cover radii up to
$10^{5}~GM/c^{2}$.

The ``diskline'' model parameters include the line energy (again fixed
at $E=6.40$~keV in all fits), the line emissivity $q$ ($J\propto
r^{-q}$), the inner radius at which the line is produced (in units of
$GM/c^{2}$, free in all fits), the outer radius at which the line is
produced (fixed at $10^{5}~GM/c^{2}$ in all fits, and unconstrained by
the data when allowed to vary), the inclination at which the emission
region is viewed, and the line flux (in units of ${\rm ph}~ {\rm
  cm}^{-2}~ {\rm s}^{-1}$).  We separately considered fits with the
line emissivity frozen at $q=3$, as per a flat accretion disk and an
isotropic source (e.g., Reynolds \& Nowak 2003, Miller 2007), and fits
with the emissivity bounded in the $2 \leq q \leq 4$ range (if the
line were to arise in a warp or wind that provides a subtantial
increase in area per radius, it is possible that the emissivity index
could be locally flatter than $q=3$).  In all fits with ``diskline'',
it is shifted using ``zmshift'' as the line model does not include a
redshift parameter.

The diskline models provide a significant improvement over the
broadened Gaussian fits, and an enormous improvement over the narrow
Gaussian fits, particularly when the emissivity is allowed to vary
(see Table 2).  In each of these fits, the emissivity drifts to
flatter values, close to $q=2$.  Relative to a broad Gaussian, the
diskline model with $q$ free represents an improvement significant at
the $4.9\sigma$, $4.6\sigma$, and $3.8\sigma$ level of confidence in
the summed, ``high/unobscured'', and ``low/obscured'' spectra
(respectively).

In the models with $q=3$ fixed, the best-fit radii are measured to be
$R_{in} = 830^{+280}_{-120}~GM/c^{2}$, $R_{in} =
590^{+190}_{-110}~GM/c^{2}$, $R_{in} = 2100^{+2700}_{-870}~GM/c^{2}$
in the summed, ``high/unobscured'', and ``low/obscured'' spectra
(respectively).  This nominally indicates that the line is produced
closer to the black hole in the ``high/unobscured'' state than in the
``low/obscured'' state.  The measured radii are smaller than the
optical BLR, but of the same order of magnitude, and these fits make
it appealing to associate the Fe~K$\alpha$ line in NGC 4151 with the
innermost extent of the optical BLR (or, an in inner X-ray portion of
a BLR that might now be viewed as spanning a larger range of radii and
wavelengths).

The best ``diskline'' fits -- those where the emissivity $q$ is
allowed to vary -- measure much smaller radii, consistent with $R_{in}
\simeq 50~GM/c^{2}$ in the summed and ``high/unobscured'' spectra and
$R_{in} \simeq 140~GM/c^{2}$ in the ``low/obscured'' spectrum.  Here
again, the radius appears to nominally be larger when the incident
flux is lower.  These fits cannot be dismissed out of hand, but the
small radii that emerge demand additional scrutiny.  For instance, the
emissivity parameter -- alone or in concert with other parameters --
might partly account for an effect other than dynamical broadening or
geometrical changes, such as scattering.  

\subsection{Modeling with ``mytorus''}
In order to address the disparate results of the fits made with
``diskline'', we next modeled the data using ``mytorus'' (Murphy \&
Yaqoob 2009, Yaqoob \& Murphy 2010).  This model includes the effects
of scattering in the nearly-Compton thick and Compton-thick media in
which Fe~K$\alpha$ lines can originate.  In general, the anticipated
line shape is {\it not} symmetric, but instead has structure to the
red of 6.40~keV owing to downscattering.  Under the right
circumstances, a secondary peak at 6.25~keV is predicted based on the
150~eV energy loss in full 180-degree Compton scattering.  Indeed, the
line shape is vaguely similar to a ``diskline'' profile.  It is
possible, then, that the shape fit only through dynamical broadening
using ``diskline'' is actually only an artifact of scattering, and
fits with ``mytorus'' represent a clear test.

In all cases, we used a stand-alone ``mytorus'' model for the line,
and a separate power-law function (e.g., ``mytorus+zpow'' in XSPEC).
Specifically, we made fits with ``mytl\_V000010nEp000H100\_v00.fits'';
like all models in the ``mytorus'' family, this is a ``table'' model,
consisting of many specific models with fixed parameters.  Best-fit
parameters can be estimated because XSPEC can interpolate between the
grid points.  This particular model was constructed assuming a
power-law source of irradiation with a terminal energy of 100~keV.

The parameters of the ``mytorus'' model include the column density of
the emitting region ($N_{\rm H}$), the inclination of the emitting
region ($\theta$), the photon index of the irradiating spectrum
($\Gamma$), the redshift of the source, and a flux normalization.  In
all of the spectra that we considered, we found that the data were
unable to constrain the column density directly, and that the measured
parameter values were insensitive to columns fixed at $N_{H} =
10^{23}~{\rm cm}^{-2}$, $N_{H} = 10^{24}~{\rm cm}^{-2}$, and $N_{H} =
10^{25}~{\rm cm}^{-2}$.  We therefore fixed a value of $N_{H} =
10^{24}~{\rm cm}^{-2}$ in all fits.  For simplicity and
self-consistency, the power-law index within the ``mytorus'' model was
tied to the value determined by the power-law continuum.  In this
initial application of ``mytorus'' without dynamical blurring, the
inclination was fixed at a value of $\theta = 15^{\circ}$, guided by
the results obtained with ``diskline''; however, the fits are largely
insensitive to this parameter.

Fits with the ``mytorus'' line function on its own did not yield
acceptable results (see Table 2).  In the summed, ``high/unobscured'',
and ``low/obscured'' spectra, the goodness of fit statistic
($\chi^{2}/\nu$) is always greater than 1.6.  The results are markedly
worse than fits with the ``diskline'' model with the emissivity fixed
at $q=3$, and dramatically worse than the ``diskline'' model where the
emissivity was allowed to vary within $2\leq q\leq 4$.  This clearly
signals that the line is unlikely to be shaped exclusively by
scattering effects, but rather by a combination of scattering and
dynamics.

We therefore proceeded to explore fits with the ``rdblur'' convolution
function acting on ``mytorus''.  The ``rdblur'' function is just the
kernel of the ``diskline'' model (e.g., when ``rdblur'' is applied to
a Gaussian function, a ``diskline'' profile is produced).  In applying
``rdblur'', the inner blurring radius was allowed to vary freely, the
outer blurring radius was again fixed at $10^{5}~GM/c^{2}$, and the
inclination of the line region was allowed to vary freely.  For
self-consistency, the inclination parameter within ``mytorus'' was
tied to the same parameter in ``rdblur''.  Two sets of fits were
again explored: one with the emissivity fixed at $q=3$, and the second
with the emissivity bounded in the range $2\leq q\leq 4$.

Comparing the results listed in Table 2, only marginally better fits
are achieved when the emissivity varies.  Relative to the results
achieved with $q=3$ fixed, smaller radii are again found, but the
small improvement in the fit statistic ($\Delta \chi^{2} = 1-2$)
signals that the small radii are not required.  The radii derived from
the fits with $q=3$ fixed are likely a more conservative set of
measurements.  In the summed, ``high/unobscured'' and ``low/obscured''
spectra, inner radii of $R_{in} = 1200^{+1400}_{-300}~GM/c^{2}$,
$R_{in} = 890^{+480}_{-220}~ GM/c^{2}$, and $R_{in} =
9200^{+90,400}_{-5700}~ GM/c^{2}$ are measured (respectively).
Importantly, these fits mark improvements over those with broad
Gaussian models that are significant at the $5.3\sigma$, $4.5\sigma$,
and $3\sigma$ level, respectively.  Although dynamical broadening
appears to be robust in the summed and ``high/unobscured'' spectra,
the evidence is more marginal in the ``low/unobscured'' phase.
Indeed, in the ``low/obscured'' phase, the line emission region is
largely unconstrained.  This is consistent with the smaller
distortions expected at the larger radii.

In this and other fits, the uncertainties in the line emission radii
are driven by the shape of the line itself and the flux uncertainties
in the line, not by uncertainties in the continuum.  Fits with the
power-law index fixed to its extremal values ($\Gamma = 1.4$, $\Gamma
= 2.0$) measure radii within errors of the best-fit measurements in
Table 2.  This holds both for fits with the emissivity fixed at $q=3$,
and fits with the emissivity bound within $2 \leq q \leq 4$.

\subsection{Reflection modeling}
As a final check, we made fits with models that jointly describe the
Fe~K$\alpha$ line and local contiuum.  We considered ``pexmon''
(Nandra et al.\ 2007) and ``xillver'' (Garcia et al.\ 2013) alone, and
then blurred by dynamical effects (``xillver'' is then known as
``relxill'').  The results of these fits are detailed in Table 3.

Both families of models have strengths and weaknesses for this
analysis.  The ``pexmon'' model includes Fe and Ni K$\alpha$ and
K$\beta$ lines, but it only describes reflection from neutral gas.  It
can be blurred with any function, including the ``rdblur'' function
used above.  In contrast, ``xillver'' can handle a range of
ionizations and self-consistently includes changes in charge state and
line breadth for different ioniations.  However, when the blurred
``relxill'' variant is used, blurring is only possible for $R\leq
990~GM/c^{2}$.  We therefore fit ``relxill'' to the summed and
``high/unobscured'' phase spectra, but we fit ``rdblur*xillver'' to
the ``low/obscured'' phase spectrum (where prior modeling suggests
larger radii).

A number of parameters are common between the ``pexmon'' and
``xillver'' models, including: the power-law index (again bounded to
$1.4\leq \Gamma\leq 2.0$), the power-law cut-off energy (fixed at
100~keV in all fits, as per the values reported in Fabian et
al.\ 2015), the ``reflection fraction'' ($f$, nominally tracing the
relative importance of the direct and reflected emission), the iron
abundance relative to solar ($A_{\rm Fe}$, fixed at unity in all
fits), the inclination at which the reflector is viewed ($\theta$),
and a flux normalization.  ``Xillver'' also measures the ionization of
the reflector ($\xi = L/nr^{2}$, where $n$ is the hydrogen number
density).  The ``relxill'' variant of ``xillver'' can measure
parameters that a model like ``rdblur*pexmon'' or ``rdblur*xillver''
cannot, including the black hole spin parameter $a$, and separate
emissivity indices for inner and outer radii.  In all fits, we fixed
the spin at a nominal value of $a=0.97$ (broadly consistent with both
Keck et al.\ 2015 and Beuchert et al.\ 2017) and tied the outer
emissivity index to the inner emissivity index, effectively just
giving a single value as per ``rdblur''.

Table 3 lists the results of several different fits, following the
pattern of Table 2.  Simple fits with ``pexmon'' and ``xillver'' -- in
the absence of dynamical blurring -- do not provide acceptable fits to
any of the spectra.  As expected, ``relxill'' provides the best
overall fits to the summed and ``high/unobscured'' spectra; the
improvement over the raw reflection models is signifcant at more than
$8\sigma$ and $6.4\sigma$, respectively.  The improvements over fits
with separate broad Gaussian functions in Table 2 are significant at
the $5.3\sigma$ and $4.5\sigma$ level of confidence, respectively.
Interestingly, dynamical blurring represents an improvement over the
best raw reflection model at the $5.6\sigma$ level of confidence in the
``low/obscured'' spectrum.  However, this model is only preferred over
the broadened Gaussian model in Table 2 at the $3.0\sigma$ level of
confidence, again signaling that Doppler broadening is likely weaker
in the ``low/obscured'' phase.

The radii measured in fits made using ``pexmon'' with the emissivity
frozen at $q=3$ are formally consistent with those where the
emissivity was merely bounded.  In the summed and ``high/unobscured''
spectra, fits with ``relxill'' measure radii that are about $\sim2$
times smaller than the radii measured using ``rdblur*pexmon''.
Moreover, ``relxill'' models are improvements over the ``pexmon''
models, preferred at the $3\sigma$ level of confidence.  In rough
terms, reflection modeling of the ``summed'' and ``high/unobscured''
spectra suggests an inner emission radius of $R_{in} =
500-1000~GM/c^{2}$.  The $1\sigma$ errors on the radii determined with
``relxill'' extend down to $R\simeq 200-300~GM/c^{2}$; this is explored in
more detail in the next section.
  
Reflection modeling of the ``low/obscured'' phase also confirms the
results detailed in Table 2.  The best overall model consists of a
blurred ``pexmon'' component, measuring $R_{in} =
3400^{+96,600}_{-2200}~GM/c^{2}$.  Note that the upper limit of the
line production region hits our upper bound, and is therefore formally
unconstrained.  This model is statistically superior to the
``xillver'' model that gives a radius of $R_{in} =
700^{+500}_{-200}~GM/c^{2}$.

In short, reflection models again point to a narrow
Fe~K$\alpha$ line production region within an inner radius of $R =
500-1000~ GM/c^{2}$.  Radii as small as $R\simeq 200~ GM/c^{2}$ are
within the $1\sigma$ confidence region in the ``high/unobscured''
state spectrum.  Particularly when the data are fit with ``relxill''
or ``xillver'', which allow for ionization effects, evidence that the
line is produced at smaller radii in the ``high/unobscured'' state
than in the ``low/obscured'' state is even weaker.  In all cases, very
low inclinations are again preferred (see Table 3).


\subsection{Difference spectroscopy}
The models considered above strongly suggest that the narrow
Fe~K$\alpha$ line in NGC 4151 is more asymmetric in the
``high/unobscured'' state than in the ``low/unobscured'' state, with a
stonger red wing indicative of emission from smaller radii.  However,
these results are necessarily model-dependent.  Difference spectra
represent a more robust and model-independent measure of the
significance at which component properties may vary.  Difference spectra
are made by subtracting the spectrum of low flux intervals from the
spectrum of high flux intervals (see, e.g., Vaughan \& Fabian 2004).
The exposure time in such intervals may not be equivalent, so it is
crucial to perform the subtraction in units of count rate, not total
counts.

XSPEC reads the exposure time of both source and background files, and
the background subtraction is performed in units of count rate, as
required.  We created a ``high/unobscured'' minus ``low/obscured''
difference spectrum by loading the summed ``high/unobscured'' spectrum
into XSPEC, and then loading the summed ``low/obscured'' spectrum
as a background.  Prior to subtraction, the spectra were grouped to
require 20 counts per bin.  In order to better understand the nature
of the variable continuum, fits were expanded to the 5.0--7.5 keV band.

The ``high/unobscured'' minus ``low/obscured'' difference spectrum is
shown in Figure 4, in the frame of the host galaxy.  The variable
continuum is consistent with a simple power-law, with $\Gamma =
1.76(7)$.  Two emission features are evident in the vicinity of the
narrow Fe~K$\alpha$ line.  When fit with simple Gaussian functions,
they are measured to have centroid energies of $E = 6.373(6)$~keV and
$E = 6.209(7)$~keV.  This is significant because values of 6.40~keV
(as per emission from neutral Fe) and 6.25~keV (as per 180-degree
Compton scattering of the 6.40~keV line) are excluded.  Dividing the
flux of the Gaussian models by their $1\sigma$ minus-side errors
suggests the featues are significant at the $4\sigma$ and $5\sigma$
level of confidence, respectively.  This represents strong,
model-independent evidence that the the red wing of the narrow
Fe~K$\alpha$ is genuinely enhanced in the ``high/unobscured'' state.

More oblique Compton scattering of the Fe~K$\alpha$ line at 6.40~keV
can lead to energy losses of less than 150~eV; this will fill-in some
flux between the line core and the 180-degree scattering limit at
6.25~keV.  However, it is not possible to generate a peak at lower
energies in a single scattering.  The tightly constrained energy of
the feature at $E = 6.209(7)$~keV would require multiple scatterings,
and a level of fine tuning that is implausible.  It is possible, then,
that the two features are Doppler-shifted emission horns from a narrow
range of radii.  The two features can be fit jointly with either a
``diskline'' component, or a ``mytorus'' component modified by
``rdblur''.  In all of these fits, an inner radius of $R =
70-80~GM/c^{2}$ is preferred, and the outer line production radius
cannot be more than 1.5 times this value (a larger run of radii would
fill-in the line profile and fail to produce the two horns).  Another
common feature of these fits is a low inclination, $\theta = 8-9$
degrees.  These models do not imply that the narrow Fe~K$\alpha$ line
in NGC 4151 is produced within such a narrow range of radii; rather,
fits to the {\it difference} spectrum merely indicate that this range
of radii contributes more strongly within the ``high/unobscured''
state than within the ``low/obscured'' state. The full line production
region is likely to span a much greater range of radii.


\subsection{Rapid line variations}
Using ``dmcopy'', we created light curves of the dispersed first-order
events from the three ObsIDs that were obtained in the
``high/unobscured'' state: 03480, 03052, and 07830.  Figure 5 plots
the light curve from each observation, as a ratio to the mean count
rate, separately in 1~ks and 5~ks bins.  In this way, the relative
amplitude of the flux variations is visible.  It is immediately
apparent in Figure 5 that 10--20\% variations of roughly 20~ks
durations are common in the ``high/unobscured'' state.  The
Fe~K$\alpha$ line is excited by the hard X-ray continuum, so line flux
variations on similar or shorter time scales to continuum variations
would independently constrain the innermost extent of the line
production region.  

The variability in Figure 5 appears to be quasi-periodic, but such
signals are rare in AGN (see, e.g, Gierlinski et al.\ 2008, Reis et
al.\ 2012), and many cycles must be observed before an apparent QPO is
significant.  To assess the observed variability, we made power
spectra from the light curves in Figure 5.  A break is evident at a
frequency just below $\nu = 10^{-4}$~Hz.  However, the addition of a
break in fits to the power spectrum is only significant at just over
the $2\sigma$ level of confidence.  Periodic and quasi-periodic
signals can be useful but they are not required to find evidence of a
reverberation signal (see, e.g., Zoghbi et al.\ 2012).

Simply dividing the events in each observation into intervals above
and below the mean count rate did not reveal variations in the narrow
Fe~K$\alpha$ line.  It is likely that events from intervals with count
rates close to the mean served to dampen any response of the line to
continuum variations.  Examination of the light curves in Figure 4
suggested that selecting events more than $\pm4$\% above and below the
mean would effectively sample the crests and troughs of the
variations, remove intervals close to the mean, and still obtain
enough total photons to permit spectral fitting.  The mean time
between the midpoint of intervals more than 4\% above and below the
mean, and the next such crest or trough, is $\Delta t = 23.0$~ks.
Assuming a black hole mass of $M_{BH} = 3.6\times 10^{7} M_{\odot}$
(Bentz \& Katz 2015), light can travel $R\simeq 130~GM/c^{2}$ in this
time.  This is close to the smallest radii indicated in some direct
spectral fits to the summed ``high/unobscured'' spectrum (see Table 2
and Table 3).

The CIAO tool ``dmgti'' was used to create corresponding good time
interval (GTI) files to extract intervals from the crests and troughs
in the 1~ks light curves.  Spectra from the crests and troughs within
each observation were then created, and corresponding response files
were generated, according to the prescription detailed previously.
The spectra from each observation were then added using the CIAO
script ``add\_grating\_spectra'', resulting in separate spectra from
all of the troughs and crests in the ``high/unobscured'' state.

Prior to fitting, the spectra were grouped to a minimum of 50 counts
per bin, in order to achieve the sensitivity needed for a clear test.
The narrow Fe~K$\alpha$ line is readily evident in the spectra from
the crests and troughs, and direct fits with various models suggested
that the line flux is higher in the crests, compared to the troughs.
When the line in each phase is fit with a broadened Gaussian or
``diskline'' model, for instance, the errors on the line flux exclude
the value in the opposing interval at the $3\sigma$ level of
confidence.

Figure 6 shows the ``crests - troughs'' difference spectrum.  In order
to achieve a good characterization of the continuum in this regime, we
expanded our fits to cover the 4--9~keV band.  It is natural that the
sensitivity of the spectrum is modest, but the narrow Fe~K$\alpha$
line at 6.4~keV is clearly evident.  Fits with a simple Gaussian
suggest that the variable part of the narrow Fe~K$\alpha$ line may be
slightly broader than in the time-averaged spectra characterized in
Table 2: $\sigma = 55^{+42}_{-22}$~eV.  Fits with a
``diskline+powerlaw'' model (zmshift*disklike+zpow, in XSPEC) also
achieve an acceptable fit.  For simplicity, we froze the emissivity at
the expected $q=3$, and $R_{in} = 130~GM/c^{2}$.  The line flux is
measured to be $K = 9(3) \times 10^{-5}~{\rm photons}~ {\rm cm}^{-2}~
{\rm s}^{-1}$, suggesting that the line is significant at the
$3\sigma$ level of confidence.  The same significance results from
fits with Gaussian models.  This level of significance is modest, but
it is model-independent, and obtained in a difference spectrum, so the
line (variation) is likely robust.

\subsection{Line imaging}
The Fe~K$\alpha$ line observed in the HETG spectra of NGC 4151 is only
consistent with low charge states, signaling that it must arise in
cold, dense gas.  The variations observed in the line properties also
strongly suggest that the line is produced in the inner accretion
flow.  As noted above, {\it Chandra} is able to resolve the optical
NLR in NGC 4151.  Wang et al.\ (2011) found that the Fe K$\alpha$
emission line is strongly dominated by nuclear emission, rather than
extended emission in the NLR (see Figure 4 in that work).  Moreover,
the diffuse X-ray emission within the NLR is consistent with
He-like and H-like charge states of O, Ne, and Mg, which is again
inconsistent with the low ionization states compatible with the
narrow Fe~K$\alpha$ emission line (Fe I-XVII).

Miller et al.\ (2017) constrained the size of the Fe K$\alpha$
emission region in NGC 1275 using a combination of sub-pixel image
processing, and ``grade=0'' event filtering.  The latter step acts to
mitigate photon pile-up distortions to the PSF by only accepting
photons that liberated electrons within a single CCD pixel (pile-up is
more likely to lead to charge in multiple pixels).  We downloaded
ObsID 9217, the longest (a net exposure of 117~ks was obtained
starting on 2008 MArch 29 at 02:53:12 UT) ACIS-S imaging observation
of NGC 4151, and reduced the data as per Miller et al.\ (2017).  The
native pixel size of 0.492 arc seconds (per side) was binned to 0.0492
arc seconds.  The event list was filtered to only accept ``grade=0''
events.  Separate event lists were created for the O VII-VIII band
(0.50-0.80 keV), Ne IX-X band (0.8-1.1 keV), and Fe K band (defined as
6.3--6.5~keV).  The images in each band were then smoothed by 6, 6,
and 3 sub-pixels respectively, in order to best represent the size and
intensity of the emission, and then combined into ``true color''
image.

The resultant image is shown in Figure 7.  Tests with the source
detection tool ``wavdetect'' suggest that the Fe K emission region is
primarily contained within a radius of about 0.3 arc seconds,
consistent with a point source and nuclear emission.  In contrast, the
O VII-VIII and Ne IX-X emission regions are several times larger.  The
sensitivity of the extended emission is not as high as in the more
detailed analysis undertaken by Wang et al.\ (2011) owing to the use
of ``grade=0'' events, but this filtering allows the nuclear and
extended emission to be disentangled.

\section{Discussion}
We have analyzed deep {\it Chandra}/HETG spectra of the Seyfert-1 AGN,
NGC 4151. We find that the narrow Fe~K$\alpha$ emission line --
sometimes associated with the torus or outer optical BLR in Seyferts
-- is asymmetric and likely shaped at least partially by Doppler
shifts and weak gravitational red-shifts.  Many of the models that we
considered suggest that the line originates approximately $R\simeq
500-1000~ GM/c^{2}$; this region is already closer to the black hole
than the optical BLR, and may be tied to a kind of X-ray broad line
region (or, XBLR).  If so, it is the first time that dynamical
imprints have clearly revealed this region in X-rays.  Some spectral
fits suggest that the line could arise at radii as small as $R\simeq
50-100~ GM/c^{2}$; this appears to be weakly confirmed by independent
variability signatures.  Changes in the shape (and, production radius)
of the line between ``high/unobscured'' and ``low/obscured'' phases,
and on shorter times within the ``high/unobscured'' state, pose some
challenges.  In this section, we review the strengths and limitations
of our analysis and results, discuss how our work impacts our
understanding of NGC 4151 and Seyferts in general, and explore the
potential for future observations to build on our work.

\subsection{Central results}
%
%
We made fits to the narrow Fe~K$\alpha$ emission line in NGC 4151
using many different models (see Table 2 and Table 3).  Simple fits
with Gaussians did not fit the data well.  Similarly, fits with models
that attempt to account for scattering in dense media and/or
ionization effects also failed to deliver acceptable fits.  In all
cases, some degree of dynamical broadening is required by the data.

Fits with the ``diskline'' model (Fabian et al.\ 1989) suggest that
the narrow Fe~K$\alpha$ line may originate as close as $R =
50~GM/c^{2}$ from the black hole, or potentially even closer.  When
the line is fit with a ``mytorus'' function (Murphy \& Yaqoob 2009,
Yaqoob \& Murphy 2010) that includes distortions to the line shape
from scattering in dense media, the required dynamical broadening is
consistent with $R\simeq 1000~GM/c^{2}$ in the summed and
``high/unobscured'' phase data when the emissivity is fixed at $q=3$
as per simple expectations far from a black hole (the radius is
largely unconstrained in the ``low/unobscured'' phase).  Separate fits
with flatter emissivity profiles return smaller radii but do not
represent statistically significant improvements.  Fits to the summed
and ``high/unobscured'' spectra with a dynamically broadened
``pexmon'' reflection model (Nandra et al.\ 2007) also measure radii
consistent with $R\simeq 1000~M/c^{2}$.  However, fits with the
``relxill'' model (Garcia et al.\ 2013) are statistically superior and
require smaller radii consistent with $R\simeq 500~ GM/c^{2}$, and
much smaller radii are not strongly excluded.  In short, radii as
small as $R = 50-100~GM/c^{2}$ are allowed by the data but are only
marginally preferred over radii of $R = 500-1000~ GM/c^{2}$.

As noted in Section 1, Bentz et al.\ (2007) used optical
BLR reverberation to derive the mass of the central black hole in NGC
4151.  Variations in the H$\beta$ line are found to lag variations in
the 5100\AA~ continuum by $\tau = 6.6^{+1.1}_{-0.8}$ days.  The size
of the H$\beta$ region can then be estimated in units of gravitational
radii via $c\tau / R_{g}$, where $R_{g} = GM_{BH}/c^{2}$.  This
simplistic estimate gives $R_{H\beta} = 3200^{+600}_{-400}~R_{g}$ for
$M_{BH} = 3.6\times 10^{7}~ M_{\odot}$.  Our results clearly suggest
that the narrow Fe~K$\alpha$ line in NGC 4151 originates at a radius
that is at least a factor of a few -- and possibly an order of
magnitude -- smaller than the optical BLR (at least in the
``high/unobscured'' state).  The narrow Fe~K$\alpha$ line might be
regarded as originating in a distinct XBLR; however, it is already
clear that the BLR is a complex, stratified goemetry, and the narrow
Fe~K$\alpha$ line could also be regarded as simply arising in its
innermost extent.

In all of the fits that we made to the narrow Fe~K$\alpha$ line in NGC
4151, a low inclination is strongly preferred in a statistical sense.
The best-fit ``diskline'' and dynamically-broadened ``mytorus'' models
for the ``high/unobscured'' state spectrum both give $\theta =
9(1)$~degrees (see Table 2).  The best-fit reflection models for the
``high/unobscured'' spectrum measure $\theta = 3(3)$~degrees and
$\theta = 8(2)$~degrees (see Table 3).  Error bars are much larger in
fits to the ``low/unobscured'' state but values are again small, and
broadly consistent with the results obtained in the more sensitive
spectra.

The inclination of the intermediate disk, then, appears to be more
consistent with values obtained via reflection studies of the
innermost disk (see, e.g., Keck et al.\ 2015, Beuchert et al.\ 2017;
also see Cackett et al.\ 2014), than with the inclination of the
optical NLR ($\theta = 45\pm5$~degrees; Das et al.\ 2005).  It is
possible that a warp in the accretion disk -- potentially tied to the
BLR -- could be the origin of this discrepancy.  In this case, the
region of the BLR that is traced by the narrow Fe~K$\alpha$ line would
still be aligned with the innermost disk (perhaps anchored into the
black hole spin plane by GRMHD effects), and the warp must occur
further out in the BLR or outer disk.  The warp would likely have to
be more extreme than the one detected in NGC 4258, which only deviates
from the midplane by $\Delta\theta = 4$~degrees (Moran et al.\ 1995).

Warps can be excited through radiation pressure (e.g., Maloney et
al.\ 1996).  However, warps and other asymmetries in AGN accretion
disks, including orbital eccentricities and spiral arm structures, can
be excited by the tidal effects of a binary companion or the orbital
passage of a massive stellar cluster (e.g., Chakrabarti \& Wiita 1993,
1994).  It is at least remotely possible, then, that the observed
inclination discrepancy is excited by another black hole or a
massive stellar cluster.  There is currently no evidence of a binary
black hole system in NGC 4151, but the system has many pecularities.
Future monitoring of NGC 4151 in multiple bands, and eventually with
high-resolution X-ray spectroscopy, can help to definitively rule out
a binary black hole system.  It may be as interesting -- and more
productive -- to explore if the passage of a massive stellar cluster
can be detected or rejected.

Time-averaged spectroscopy and reverberation results strongly
suggest that the accretion disk extends close to the ISCO in NGC 4151
(e.g., Zoghbi et al.\ 2012, Keck et al.\ 2015, Beuchert et al.\ 2017).
The fact of a narrow Fe~K$\alpha$ line with mild relativistic shaping
merely points to an intermediate disk structure.  Our modeling is
insensitive to reflection from the innermost accretion disk, and our
results do not imply that the accretion disk is truncated at
intermediate radii.

\subsection{Long term variability and radius variations}
Lines excited by radiation from central engine are expected to follow
$R \propto L^{1/2}$.  This trend is certainly observed in studies of
the optical continuum and high-ionization optical lines linked to the
BLR, including H$\beta$ in NGC 4151 (e.g., Bentz et al.\ 2013).
Indeed, the tight relationship between AGN luminosity and BLR size
suggests that the BLR may originate in the disk itself and generate a
failed wind.  Some of our fits nominally suggest an {\it inverse}
relationship between the X-ray continuum and the radius at which the
narrow Fe~K$\alpha$ line is produced.  If this line is tied to the
innermost extent of the BLR or XBLR, it would suggest that the inner
radius is changing, or that the radius at which the line is detectable
is changing in an unexpected manner.  It is therefore worth critically
examining the data, and exploring possible explanations.

Evidence of dynamical broadening of the Fe~K$\alpha$ line is strongest
in the summed spectrum, and in the ``high/unobscured'' state; the
evidence is weaker in the ``low/obscured'' phase.  Accordingly, direct
spectral fits to the ``low/obscured'' state with various models yield
correspondingly larger errors on the line production region (see
Tables 2 and 3).  However, the ``high/unobscured'' minus
``low/obscured'' {\it difference} spectrum makes clear that the narrow
Fe~K$\alpha$ line is more skewed -- and originates at smaller radii --
in the ``high/unobscured'' state (see Section 3.7 and Figure 4).  Fits
to this difference spectrum nominally indicate that a fairly narrow
range of radii may be emphasized in the ``high/unobscured'' state.
This could indicate changes in the local disk structure that enhance
reflection; warps and clumpy, failed winds may be viable explanations.
Structures of this sort might help to explain evidence of a second
reprocessing in NGC 4151, based on UV and X-ray monitoring (Edelson et
al.\ 2017).

The physical processes that could underpin such explanations may reach
to the nature and origin of the BLR:

High ionization optical lines in the BLR are likely produced at a
greater height above the disk than lower-ionization optical lines;
this is consistent with a wind that has had the chance to flow some
distance (e.g., Collin-Souffrin et al.\ 1998, Kolatchny et al.\ 2003;
also see Czery et al.\ 2016).  The relatively simple and unobscured
path for radiation between the central engine and optical BLR high
above the disk means that $R\propto L^{1/2}$ can play out.  This
expected radius--luminosity relationship implicitly assumes that the
geometry of the irradiated gas, and relevant optical depths, do not
change in response to flux variations from the central engine.  This
assumption may not be valid for the narrow Fe~K$\alpha$ line in some
Seyferts.

At least in NGC 4151, the narrow Fe~K$\alpha$ line now appears to
originate at smaller radii than the optical BLR, though it traces cold
gas (where dust may be present).  It has been suggested that the BLR
may be exist at least partially owing to the influence of dust: the
higher cross section of dust relative to gas may be the key to lifting
material above the disk (see, e.g., Czerny et al.\ 2011, Czerny et
al.\ 2016).  It is possible, then, that the narrow Fe~K$\alpha$ traces
the region where dust and gas are initially lifted upward.  The
enhanced local solid angle creted through this process might mimic a
warp, but it is also the case that dust can be an important factor in
creating actual warps within disks (Maloney et al.\ 1996).  Enhanced
radiation from the central engine might eventually destroy dust within
a larger radius, reducing contributions to the Fe~K$\alpha$ line from
small radii.  However, the enhanced local solid angle might also
shield larger radii from the central engine and serve to limit
narrower contributions to the Fe~K$\alpha$ emission line.

At least one aspect of this explanation can be explored
quantitatively.  Czerny et al. (2016) predict the dust sublimation
radius -- possibly the effective inner edge of the BLR -- as a
function of black hole mass and Eddington fraction.  Based on UV and
X-ray data, Crenshaw et al.\ (2015) report that NGC 4151 has an
average bolometric luminosity of $L_{bol} = 7.4\times 10^{43}~ {\rm
  erg}~ {\rm s}^{-1}$.  Assuming a mass of $3.6\times
10^{7}~M_{\odot}$ (Bentz \& Katz 2015), this equates to an Eddington
fraction of $L_{bol}/L_{Edd} \simeq 0.016$.  The work by Czerny et
al.\ (2016) then predicts a dust sublimation radius of $R_{dust}
\simeq 1-2\times 10^{16}~ {\rm cm}$, or $R_{dust} \simeq 1900-3800~
GM/c^{2}$.  This radius is at least a factor of a larger than our
estimates based on fits to the narrow Fe~K$\alpha$ line, and possibly
an order of magnitude larger.  Given the numerous uncertainties in the
input parameters, however, the prediction might be regarded as broadly
consistent with our results.  Further development of the Czerny et
al.\ (2016) model incorporating explicit consideration of Fe~K$\alpha$
line production, and more sensitive data, may be able to formally
reconcile the details.

\subsection{Mass Estimates via the Narrow Fe K$\alpha$ Line}
Within the ``high/unobscured'' state, the continuum flux is highly
variable (see Figure 5).  Selecting the ``crests'' and ``troughs'' of
the light curve (periods greater than 4\% above and below the mean),
direct fits to the separate spectra reveal variations in line flux
that are significant at the $3\sigma$ level of confidence (see Section
3.8).  The ``crests-troughs'' difference spectrum also reveals an
Fe~K$\alpha$ line significant at the $3\sigma$ level of confidence
(see Figure 6).  This signals that the line is responding to the
continuum variations on the characteristic time scale of the
variability ($\Delta t = 23.0$~ks), independently indicating that
inner extent of the line production region may be as small as $R\simeq
130~GM/c^{2}$, or even smaller.

Potentially coupled variability in the X-ray continuum and narrow
Fe~K$\alpha$ line, combined with the ability measure subtle dynamical
shaping of the line, opens the possibility of adapting optical BLR
reverberation techniques and measuring black hole masses in X-rays.
Whether the narrow Fe~K$\alpha$ line originates in the disk, in a
wind, or in a combination of these that marks the innermost extent of
the BLR, the local gas motions are likely to be largely Keplerian.  In
this case, the black hole mass can be estimated by:

\[M_{BH} \propto rv^{2}/G\]

\noindent where $M_{BH}$ is the mass of the black hole, $r$ is the
radius (in physical units) where the line is produced, $v$ is the {\it
  full} velocity of the gas, and $G$ is Newton's gravitional constant.

In the limit of excellent data, fits with a relativistic line model
(or, reflection model modified by dynamical broadening) will give
tight constraints on the innermost line production radius.  These
models map the radius in units of gravitational radii, $GM/c^{2}$.
The full velocity in Keplerian orbits is then given by $v^{2}/c^{2} =
1/N$, where $N$ counts the number of gravitational radii.  Thus, the
velocity $v$ can actually be deduced through the radius in
gravitational units.  The radius $r$ must be in physical units,
however, and obtained from the characteristic variability timescale:
$r = c\Delta t$.

The overall best-fit model for the Fe~K$\alpha$ line in the variable
``high/unobscured'' phase of NGC 4151 is obtained with ``relxill''
(see Table 3), which measures an inner line production radius of $R =
520^{+10}_{-350}~ GM/c^{2}$.  Again assuming $\Delta t = 23.0$~ks, a
black hole mass of $M_{BH} = 0.9-2.6\times 10^{7}~M_{\odot}$ results.
This mass is 1.4--4.0 times lower than that derived by Bentz \& Katz
(2015) using optical reverberation techniques, but the difference is
comparable o the scatter of the $M-\sigma$ relationship (0.46 dex,
Gultekin et al.\ 2009).  Higher mass estimates result from smaller
estimates of the Fe~K$\alpha$ line emission radius.  Some aspects of
our analysis point to smaller radii, especially within the
``high/unobscured'' state; additional data and improved modeling may
strengthen these hints and could potentially result in higher mass
estimates.

Estimates of this sort are at the limit of the {\it Chandra} data that
have been obtained so far.  The sensitivity of these data curtails the
quality of the radius constraints that can be obtained, in part
because various plausible models are not differentiated at a high
level of statistical confidence.  More precise radii should be readily
obtained from nearby Seyferts using the calorimeter spectrometers
aboard {\it XARM} and {\it ATHENA} (see below).  However, further
observations of NGC 4151 and other Seyferts with {\it Chandra}, and
more distant and/or fainter Seyferts with {\it XARM}, may result in
data of similar quality.  In this case, it may be pragmatic to adapt
the technique to rely on simpler line models and measurements of the
inclination.
 
The {\it observed} width of the Fe~K$\alpha$ line -- as measured by a
Gaussian -- does not reflect the full velocity that is required in the
equation.  Rather, $v_{obs.} = v_{Kepl.}\times {\rm sin}(\theta)$,
where $\theta$ is the inclination of the disk relative to the line of
sight.  If the narrow Fe K$\alpha$ line at 6.40~keV is crudely fit
with a simple Gaussian function that gives the width $\sigma$ of the
line in units of eV, then the full velocity would be

\[v = (c\sigma/6400)/sin(\theta).\]

The shifts that relativistic orbital motion imprint onto line profiles
depend on the depth of the local potential, and the inclination at
which the disk is viewed.  Thus, in cases where the narrow
Fe~K$\alpha$ line is asymmetric and can constrain the parameters of a
relativistic line model, the inclination $\theta$ can be measured
directly.  In other cases, it may be more pragmatic to take a value of
$\theta$ from fits to reflection from the innermost accretion disk,
close to the ISCO.

{\it Physical} radii cannot be determined using relativistic line models;
the line shifts only depend on the radius in units of $GM/c^{2}$.
Again writing the radius in physical units via $r = c\Delta t$, where
$\Delta t$ is the characteristic timescale on which the line varies.
Then,

\[M_{BH} \propto (c\Delta t) (c\sigma/6400)^{2}/ G sin^{2}(\theta).\]

\noindent To be clear, the inclination is taken from fits to the data
with a relativistic line function (or independent fits to reflection
from the innermost disk), and the gas velocity $\sigma$ is measured
from separate fits to the data with a simple Gaussian function.

Sensible minimal requirements for estimating a black hole mass using
this method might include flux variability in the continuum and
Fe~K$\alpha$ line, and exposure durations that sample enough eposides
of the variability.  When implemented in this manner, we expect
that the errors on mass will be driven by the errors on the
inclination, $\theta$.  These errors will also be minimized in
high-resolution, calorimeter spectra.

In the case of NGC 4151, Gaussian fits to all of the summed, and
time-averaged spectra from the ``high/unobscured'' and
``low/obscured'' states are consistent with $\sigma = 23(2)$~eV.
However, it is more appropriate to use the variable line width of
$\sigma = 55^{+42}_{-22}$~eV, based on the ``crests-troughs''
difference spectrum obtained in the ``high/unobscured'' state.  Again,
the mean time beteen the ``crests'' and ``troughs'' of the light curve
is $\Delta t = 23.0$~ks.  We made numerous fits with several different
models, but the average of the centroid values derived in models
allowing for dynamical broadening in the ``high/unobscured'' state is
$\theta = 7.7$~degrees.  These numbers yield a black hole mass of $M =
0.7-6.0 \times 10^{7}~M_{\odot}$.  This estimate is formally consisent
with the optical BLR reverberation mass, but with a larger
uncertainty.  It must be noted that the estimate is very sensitive to
the inclination.

The estimates explored here should likely not be regarded as
measurements, but rather as early proofs-of-principle.  For instance,
the characteristic time scale used in these calculations is treated as
a delay time, but only a small number of variations have been
observed.  The characteristic time scale could have a different
explanation tied to accretion phenomena at smaller radii; in that
case, it may still be possible to derive masses, but the origin of the
characteristic time scale will have to be understood in terms of
thermal or viscous times.  If reverberation from the XBLR using Fe
K$\alpha$ lines proves viable, the improved sensitivity of the X-ray
calorimeter spectrometers aboard $XARM$ and {\it ATHENA} may
enable mass estimates via this means in a large set of AGN (see below).

\subsection{Comparisons to prior work}
Shu et al.\ (2010) examined narrow Fe~K$\alpha$ emission lines in a
set of Seyfert AGN observed with the {\it Chandra}/HETGS.  The summed
line profile in NGC 4151 is found to have a width of $FHWM =
2250^{+400}_{-360}$~km/s.  Our measurement of the width using Gaussian
models to the time-averaged summed spectrum produces formally
consistent results.  Via Gaussian modeling, then, the FWHM suggests a
region with a size of $R_{Fe~K\alpha} = 17800^{+7400}_{-5000}~ R_{g}$.
This is nominally several times {\it larger} than the H$\beta$ line
region inferred from via optical reverberation mapping.

Our work expands upon that undertaken by Shu et al.\ (2010) in
numerous ways.  Among these is that we considered additional
observations (ObsIDs 16089, 16090; see Table 1), we binned the data
more aggressively and employed improved weighting techniques, we
examined a broad array of non-Gaussian functions that allow for
asymmetry driven by dynamical effects and scattering, and we examined the
variability of the line in numerous ways.  Our results suggest that
the Gaussians employed by Shu et al.\ (2010) measured projected
velocities that are a small fraction of the full velocity because the
central engine is viewed at a low inclination.  Improved modeling that
accounts for inclination effects places the narrow Fe~K$\alpha$ line
production region within the optical BLR, or likely at even smaller
radii.

It is possible to utilize higher order {\it Chandra}/HETG spectra to
better understand some sources (see, e.g., Miller et al.\ 2015, 2016).
In the Fe~K band, the resolution of third-order HEG spectra is
approximately 15~eV, midway between the resolution of the first-order
{\it Chandra} HEG and that anticipated with {\it XARM}.  Liu (2016)
considered the third-order HEG spectrum of NGC 4151, in an effort to
better understand the true velocity width of the Fe~K$\alpha$ line.
Indeed, the best-fit Gaussian model for the the line in the
third-order spectrum was nominally measured to be narrower than the
line width derived from the first-order spectrum.  However, the errors
on the third-order spectrum are 2--3 times larger than in the
first-order spectrum, and the $1\sigma$ confidence intervals from the
spectra easily overlap.  In a formal statistical sense, the
measurements are consistent.  There is no evidence that the
first-order spectra has provided a false view of the narrow
Fe~K$\alpha$ emission line in NGC 4151.

The low inclinations that have recently been measured from the
innermost disk in NGC 4151, and now also from the narrow Fe~K$\alpha$
line, are notionally incongruous with obscuration that is -- at least
in some phases -- more like that observed in Seyfert-2 AGN and
attributed to an equatorial line of sight.  Although NGC 4151 is a
standard Seyfert-1 in optical, it is apparently more complex in
X-rays.  Recent work on NGC 5548 and NGC 3783 has shown that these
well-known Seyfert-1s occasionally undergo episodes of strong
obscuration (Kaastra et al.\ 2014, 2018).  These episodes only came to
light through unprecedented monitoring and joint observing campaigns,
so it is possible that enhanced obscuration in Seyfert-1s is fairly
common but has simply evaded detection. 

It is interesting to note that Shu et al.\ (2010) found two more
Seyferts wherein the narrow Fe~K$\alpha$ line is narrower than the
H$\beta$ line: NGC 5548 and NGC 3783.  The same projection effects
that we infer in NGC 4151 may be at work in these famous sources.  The
narrow Fe~K$\alpha$ emission lines in NGC 3783 and NGC 5548 are not as
bright nor as prominent as the line in NGC 4151, and this may hinder a
clear result using {\it Chandra}.

X-ray grating spectroscopy of other Seyferts has previously revealed
some evidence that the BLR may extend as close to the central black
hole as $R \simeq few\times 100~GM/c^{2}$.  A deep {\it Chandra}/LETGS
spectrum revealed evidence of broadened He-like and H-like C, N, O,
and Ne lines in Mrk 279 (Costantini et al.\ 2007), with typical
velocities of $1-2\times 10^{4}$~km/s (FWHM).  Later observations with
the {\it XMM-Newton}/RGS only found marginal evidence for broadened
lines (Costantini et al.\ 2010).  Additional evidence of X-ray lines
from $R\simeq few\times 100~GM/c^{2}$ has recently been detected in
Mrk 509 (Detmers et al.\ 2011), and NGC 3783 (Kaastra et al.\ 2018),
among others.  In each of these cases, the broadening is symmetric.

\subsection{Future studies}
The X-ray flux of NGC 4151, and the strength of its Fe K$\alpha$ line,
set it apart from other Seyferts.  A dedicated multi-cycle {\it
  Chandra}/HETGS monitoring program that systematically samples the
relevant time scales can potentially achieve the first X-ray
reverberation study of intermediate regions of the accretion disk.
Such a study could set the stage for future efforts with {\it XARM}
and {\it ATHENA}.  Although the potential of calorimeters for
reflection spectroscopy of reflection from the innermost accretion
disk has been examined, studies of intermediate disk radii and the BLR
are less developed (see, e.g., Nandra et al.\ 2013, Reynolds et
al.\ 2014).  We have therefore constructed a small set of simulations
to illustrate the potential of future calorimeter spectra.

Current versions of the ``xillver'' and ``relxill'' models (e.g.,
Garcia et al.\ 2013) are not yet suited to calorimeter data: the
energy resolution of these models in the Fe K band is approximately
18--20~eV, which is coarser than the 5~eV and 2.5~eV resolution
anticipated from {\em XARM} and {\em ATHENA} (e.g., Barret et
al.\ 2018).  The ``pexmon'' model (Nandra et al.\ 2007) includes
multiple lines at proper relative strengths, but the model does not
include complex line structure.  In contrast, ``mytorus'' (Murphy \&
Yaqoob 2009, Yaqoob \& Murphy 2010) has a resolution of 0.4~eV in the
Fe K band, and includes line structure created by scattering in dense
gas.

Whereas the advantages of ``mytorus'' are small at HETGS resolution,
they may be important at calorimeter resolution.  For this reason, we
have built simulations based on the best-fit broadened ``mytorus''
models with $q=3$ in Table 2.  We examined two of the larger innermost
line production radii indicated by the data, $R = 500~GM/c^{2}$ and
$R=1000~GM/c^{2}$, as large radii have smaller impacts on the line
asymmetry and require calorimeter resolution in a greater degree.  The
line flux was held constant as the radius was varied.  We adjusted the
internal resolution of the ``rdblur'' function from 20~eV to 0.4~eV in
order to generate the simulated spectra.  Based on the characteristic
$\Delta t\simeq20$~ks variations in NGC 4151, we created sets of 20~ks
snapshot spectra using established {\it Hitomi} (as a proxy for {\it
  XARM}) and {\it ATHENA} responses, and the ``fakeit'' function
within XSPEC.

Figure 8 shows the results of fits to these simulated spectra.  The
dynamical content of the line profiles is especially clear.  Even in
shorter exposures, strong constraints on the line production radius
would be obtained using {\it XARM}.  The sensitivity achieved in short
{\it ATHENA} exposures is tremendous; fine details should be
detectable that will allow for precise reverberation mapping of the
emission region.  More importantly, the sensitivity afforded by {\it
  ATHENA} should enable narrow Fe~K$\alpha$ emission line spectroscopy
in far more distant galaxies, greatly increasing the total number of
AGN in which such studies are possible.\\

JMM acknowledges Keith Arnuad, Xavier Barcons, Misty Bentz, Niel
Brandt, Bozena Czerny, Javier Garcia, Julian Krolik, Mike Nowak, and
Tahir Yaqoob for helpful discussions.  We thank the anonymous referee
for a careful examination of this paper that led to improvements.  JMM
is grateful to NASA for support through the {\em Astro-H} Science
Working Group.

\clearpage

\begin{table}[t]
\caption{Chandra/HETGS Observations}
\begin{footnotesize}
\begin{center}
\begin{tabular}{llll}
ObsID   &   Start Time (MJD)   &   Net Exposure (ks) & state \\
\tableline
00335   &    51609.0  & 47.4 & low/obscured\\
03480   &    52402.0  &  90.8 & high/unobs.\\
03052   &    52403.8  &  153.1 & high/unobs.\\
07829   &    54178.5  &  49.2 & low/obscured \\
07830   &    54302.4  &  49.3 & high/unobs.\\
16089  &    56700.8  &  171.9 & low/obscured \\
16090  &    56724.6  &   68.9 & low/obscured \\
\tableline
Total  &   --        &    630.6 & -- \\
\tableline
\tableline
\end{tabular}
\vspace*{\baselineskip}~\\ \end{center} 
\tablecomments{Chandra/HETGS observations of NGC 4151.  The
  observation identification number, the observation start time (in
  MJD), and the net exposure are listed.  Observations of NGC 4151
  sample two states (see Figure 1), and the state identification for
  each exposure is also listed. }
\vspace{-1.0\baselineskip}
\end{footnotesize}
\end{table}
\medskip

\begin{table}[t]
\caption{Spectral Fits with Separate Line Functions}
\begin{footnotesize}
\begin{center}
\begin{tabular}{llllllllll}
spectrum & components &  $\Gamma$           &     $K_{zpow}~(10^{-2})$ &   $\sigma$~(eV)  &  $K_{line}~(10^{-X})$ &   $R_{in}~(GM/c^{2})$  &  incl. (deg.) &   $q$   & $\chi^{2}/\nu$ \\ 
\tableline
\tableline
summed  &     zpow+zgauss  &  $1.4^{+0.3}$       &  $2.5^{+0.7}_{-0.4}$  &    0*      &     1.2(1)     &      --               &  --   &       --     &            336.3/82\\
summed  &     zpow+zgauss  &  $1.8^{+0.3}_{-0.2}$  &  $5.3^{+3.6}_{-2.1}$ &     23(2)  &     1.67(7)     &      --              &   --  &        --   &              155.3/81\\
\tableline
summed  & zpow+zms*diskline &  $1.7(3)$          &  $4.0^{+3.0}_{-1.6}$ &    --      &     1.58(7)  &     $830^{+280}_{-120}$   &  $9(2)$  &      3.0*       &           144.5/80\\
summed  & zpow+zms*diskline &  $1.4^{+0.2}$      &   $2.1(1)$         &     --     &      2.1(1)   &      $52^{+10}_{-7}$      &  $11(1)$ &      $2.13(5)$  &           109.3/79\\
\tableline
summed  & mytorus+zpow          &  $1.4^{+0.2}$      &   $2.2^{+1.1}_{-0.1}$   & --       &  $1.7^{+1.2}_{-0.6}$ &  --               &      15*  &     --          &        215.6/82 \\
summed  & rdblur*mytorus+zpow  &  $1.5^{+0.3}_{-0.1}$ &  $2.5^{+1.0}_{-0.1}$ &   --     &  $2.5^{+2.7}_{-0.6}$ & $1200^{+1400}_{-300}$ &  $12(3)$ &     3.0*        &        109.8/80 \\
summed  & rdblur*mytorus+zpow  &  $1.6(2)$         &  $2.8^{+1.8}_{-0.7}$ &   --      &  $3.3^{+2.1}_{-1.2}$ & $440^{+610}_{-190}$  &  $16(3)$ &     $2.0^{+0.2}$   &       107.3/79 \\
\tableline
\tableline
high/unobs. & zpow+zgauss  &   $1.4^{+0.4}$     &   $3.8^{+3.3}_{-0.6}$  &  0*     &       1.2(1)    &          --             &    --  &       --   &              188.2/82\\
high/unobs.  & zpow+zgauss  &  $ 1.8(3)$      &   $7.3^{+2.7}_{-3.4}$  &  23(2)  &      1.7(1)    &       --                 &   --   &      --     &             113.7/81\\
\tableline
high/unobs. & zpow+zms*diskline &  $1.7(3)$    &    $5.4^{+4.8}_{-2.6}$ &  --  &    1.63(1)   &         $590^{+190}_{-110}$   &   $8(1)$   &     3.0*        &        103.6/80\\
high/unobs. & zpow+zms*diskline &  $1.4^{+0.4}$  &   $3.1^{+1.2}_{-0.7}$  &  --  &    2.2(2)   &         $56^{+8}_{-9}$        &  $9(1)$    &    $2.2(1)$    &           86.3/79\\
\tableline
high/unobs. & mytorus+zpow        &  $1.4^{+0.2}$    &   $3.5^{+3.1}_{-1.1}$ &   --     &  $2.0^{+2.5}_{-0.3}$  & --               &    15*   &      --        &          136.5/82 \\
high/unobs. & rdblur*mytorus+zpow &  $ 1.6(2)  $   &    $4.1^{+3.6}_{-1.4}$ &   --     &  $3.1^{+3.8}_{-1.0}$ & $890^{+480}_{-220}$ &  $9(2)$  &    3.0*         &       87.9/80 \\
high/unobs. & rdblur*mytorus+zpow &  $ 1.4^{+0.3}$  &    $3.2^{+1.3}_{-1.2}$  &  --     &   $2.2(2)$        &  $58^{+9}_{-10}$    &  $9(1)$ &     $2.2^{+0.1}$  &       86.3/79 \\
\tableline
\tableline
low/obs. &     zpow+zgauss  &  $1.4^{+0.3}$      &  $1.4^{+0.9}_{-0.2}$  &   0*   &      1.2(1)    &         --                &    --  &         --   &            167.3/39\\
low/obs. &     zpow+zgauss  &  $1.7_{-0.3}^{+0.3}$ &  $2.3^{+0.4}_{-0.4}$ &   22(2)  &     1.6(1)   &         --                &  --    &       --     &            53.5/38\\
\tableline
low/obs. & zpow+zms*diskline &  $1.6^{+0.4}_{-0.2}$ & $1.8^{+2.7}_{-0.4}$  & --   &    $1.5(1)$        &    $2100^{+2700}_{-860}$ &  $15^{+8}_{-3}$ &   3.0*         &        58.4/37\\
low/obs. & zpow+zms*diskline &  $1.55^{+0.4}$      &  $3.1^{+1.2}_{-0.7}$ & --   &   $1.8^{+0.3}_{-0.1}$ &  $140^{+120}_{-50}$     &  $17(4)$      &   $2.2{+0.02}$  &        32.7/36\\
\tableline
low/obs. & mytorus+zpow         &  $1.4^{+0.2}$     &   $1.3^{+0.6}_{-0.2}$  &    --    & $1.7^{+1.1}_{-0.1}$  &  --                &   15*         &   --         &         94.1/39 \\
low/obs. & rdblur*mytorus+zpow  &  $1.4^{+0.3}$     &   $1.3^{+1.1}_{-0.2}$  &    --    & $2.0^{+2.2}_{-0.1}$  & $9200^{+90400}_{-5700}$ &  $26^{+3}_{-9}$ &  3.0*       &         41.6/37 \\
low/obs. & rdblur*mytorus+zpow  &  $1.4^{+0.3}$     &   $1.3^{+1.1}_{-0.2}$  &    --    &  $2.0^{+1.5}_{-0.1}$ & $790^{+4150}_{-460}$   &   $17^{+12}_{-3}$ & $2.0^{+0.1}$ &         39.2/36 \\
\tableline
\tableline
\end{tabular}
\vspace*{\baselineskip}~\\ \end{center} 
\tablecomments{The results of spectral fits with simple continua and
  additive line models.  Fits to the first-order HEG spectra from the
  total summed spectrum, ``high/unobscured'', and ``low/obscured''
  spectra are listed separately.  All spectral fits were local, and
  made over the 6.0-6.7~keV band.  In all cases, a redshift of
  $z=0.0033$ was used in models where the redshift could be set.  The
  summed and ``high/unobscured'' spectra were grouped to require 20
  counts per bin; the ``low/obscured'' spectra were grouped to require
  100 counts per bin (to give approximately similar sensitivity in
  each bin).  In all cases, the index of the power-law component
  (zpow) was restricted to the range $1.4\leq \Gamma \leq 2$.  Three
  different line functions are considered: simple Gaussians (zgauss),
  the ``diskline'' model, and a version of ``mytorus''.  Fits with the
  Gaussian and ``diskline'' models fixed the line centroid energy at
  6.40~keV; allowing a range of 6.40-6.43 keV (Fe I-XVII) does not
  alter the results significantly.  The ``diskline'' component does
  not include a redshift parameter and was adjusted using ``zmshift''
  (zms).  The ``mytorus'' model used does not have an independent line
  centroid paramter but includes a redshift.  We fit the model
  assuming a column density of $N_{H} = 1.0\times 10^{24}~{\rm
    cm}^{-2}$; fits with values of $10^{23}~{\rm cm}^{-2}$ and
  $10^{25}~{\rm cm}^{-2}$ did not change the measured parameters.  The
  ``mytorus'' model was considered without weak relativistic skewing,
  and separately fit after convolution with the kernel of the
  ``diskline'' function (rdblur).  Within ``diskline'' and ``rdblur''
  the emissivity index was bound in the range $2\leq q\leq 4$ (where
  $J\propto r^{-q}$).  A value of $q=3$ is expected for reflection
  from a thin disk far from the black hole; we have allowed for a
  range to account for a disk that may not be locally flat.  The
  normalizations are given in units of $10^{-X}$; $X=4$ for the
  Gaussian and ``diskline'' components; $X=2$ for the ``mytorus''
  normalizations.  Parameters marked with an asterisk were fixed at
  the value given.}
\vspace{-1.0\baselineskip}
\end{footnotesize}
\end{table}
\medskip

\begin{table}[t]
\caption{Spectral Fits with Disk Reflection Models}
\begin{footnotesize}
\begin{center}
\begin{tabular}{llllllllll}
spectrum &  components &  $\Gamma$     &  $K_{refl}~(10^{-X})$ &      $R_{in}~(GM/c^{2})$  &  incl. (deg.) &   $q$   &    $f_{refl}$  &  log~$\xi$ &   $\chi^{2}/\nu$ \\ 
\tableline
\tableline
summed    & pexmon        &    1.5(3)   &  $2.8^{+1.8}_{-1.0}$  &     --                 &  15*       &   --                 &  $0.40^{+0.05}_{-0.08}$ &   -- &             247.4/81  \\
summed    & rdblur*pexmon &    1.7(3)   &  $3.9^{+2.9}_{-1.7}$  &     $1100_{-240}^{+990}$ &  $11(3)$   &    3.0*              &  $0.57^{+0.08}_{-0.11}$ &   -- &           114.2/80    \\
summed    & rdblur*pexmon &    1.7(3)   &  $3.8^{+2.5}_{-1.6}$  &    $1150^{+750}_{-470}$  &  $12(3)$   &    $2.5^{+0.3}_{-0.5}$ &  $0.56^{+0.11}_{-0.04}$ &   -- &             113.5/79  \\
\tableline
summed     & xillver          &  $1.6(2)$           &   $7.0(1)$          &  --                   &  15*           &    --               &  0.7(1)  &  $0.3^{+0.7}_{-0.3}$ & 345.0/81 \\
summed     & relxill          &  $1.5_{-0.1}^{+0.3}$  &   $8^{+7}_{-1}$      &   $520^{+10}_{-230}$    &  $3_{-3}^{+1}$  &    $3.4_{-1.2}^{+0.6}$ &  0.8(2) &   $0.0^{+0.1}$   & 100.8/78 \\
\tableline
\tableline
high/unobs. &   pexmon        &   $1.5_{-0.1}^{+0.3}$ &  $4.0^{+3.6}_{-1.4}$ &   --                 &  15*      &   --                 &  $0.28(5)$  &   -- & 151.7/81  \\
high/unobs. &  rdblur*pexmon  &   1.7(3)            &  $5.2^{+4.7}_{-1.7}$ &   $830^{+370}_{-200}$  &  $8(2)$   &    3.0*              &  $0.38(6)$  &   -- & 90.6/80    \\
high/unobs. &  rdblur*pexmon  &   1.7(3)            &  $5.2^{+4.7}_{-1.7}$ &   $970^{+620}_{-410}$  &  $8(2)$   &    $3.4^{+0.6}_{-0.8}$ &  $0.38(7)$  &   -- & 90.4/79    \\
\tableline
high/unobs. & xillver         &   1.9(1)             &   $10.9(1)$        &   --                  &  15*         &    --                 &  $0.55_{-0.20}^{+0.05}$ & $0.15_{-0.15}^{+0.05}$ & 195.3/81 \\
high/unobs. & relxill          &  $1.6^{+0.4}_{-0.1}$  &   $11.1(2)$         &   $520^{+10}_{-350}$   &  $3(3)$      &    $3.9_{-0.8}^{+0.1}$  &  0.56(3) &  $0.68^{+0.95}_{-0.08}$ &   81.4/78 \\
\tableline
\tableline
low/obs.   & pexmon           &   $1.4^{+0.6}$        & $1.3^{+1.1}_{-0.1}$  &   --                   &     15*        &       --       &  $0.65^{+0.14}_{-0.14}$   & -- &     109.5/39  \\
low/obs.   & rdblur*pexmon    &   $1.5_{-0.1}^{+0.4}$  & $1.4^{+1.4}_{-0.2}$  &  $10000^{+39200}_{-5700}$ &    $28(10)$   &      3.0*       &  $0.9^{+0.1}_{-0.3}$   &  -- &  39.8/37   \\
low/obs.   & rdblur*pexmon    &   $1.7$              & $5.2$              &  $3400^{+96600}_{-2200}$ &    $25^{+16}_{-8}$  &  $2.0^{+0.4}$ &  $0.8^{+0.4}_{-0.1}$  &  -- &  38.5/36 \\
\tableline
low/obs.    &  xillver         &  $1.4^{+0.4}$        &  $5.4^{+0.2}_{-1.3}$  &     --               &   15*        &    --   & 1.0(1) &  $0.3_{-0.3}^{+0.8}$   &                123.5/38 \\
low/obs.    & rdblur*xillver   &  $1.4^{+0.4}$        &  $5.2^{+0.2}_{-1.4}$ &  $700^{+500}_{-200}$    &   $3(3)$       &  3*   &  $1.2(2)$ & $0.01_{-0.01}^{+0.4}$ &   46.1/37 \\
\tableline
\tableline
\end{tabular}
\vspace*{\baselineskip}~\\ \end{center} 
\tablecomments{The results of spectral fits with reflection models.
  Fits to the first-order HEG spectra from the total summed spectrum,
  ``high/unobscured'', and ``low/obscured'' spectra are listed
  separately.  All spectral fits
  were local, and made over the 6.0-6.7~keV band.  In all cases, a
  redshift of $z=0.0033$ was used in models where the redshift
  could be set.  The summed and ``high/unobscured'' spectra were
  grouped to require 20 counts per bin; the ``low/obscured'' spectra
  were grouped to require 100 counts per bin (to give approximately
  similar sensitivity in each bin).  In all cases, the index of the
  power-law component (zpow) was restricted to the range $1.4\leq
  \Gamma \leq 2$.  Within the reflection models, the high-energy
  cut-off to the power-law was fixed at 100~keV (as per Fabian et al.\ 2015),
  and the iron abundance was fixed at $A_{Fe} = 1.0$.  Within
  ``relxill'', the black hole spin parameter was fixed at $a=0.98$;
  the fits were insensitive to this parameter as the radii of interest
  are far from the black hole.  The flux normalization of the
  reflection models is given in units of $10^{-X}$; $X=4$ for the ``pexmon''
  models, and $X=2$ for the ``relxill'' models.}
\vspace{-1.0\baselineskip}
\end{footnotesize}
\end{table}
\medskip

\clearpage

\begin{figure}
  \includegraphics[scale=0.7,angle=-90]{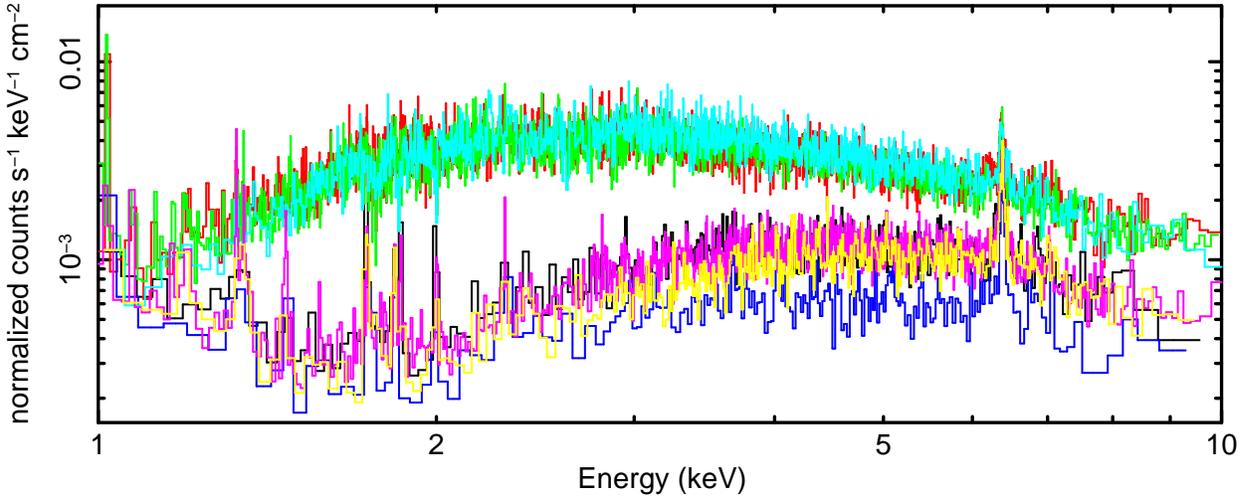}
  \figcaption[t]{\footnotesize First-order {\it Chandra}/HEG spectra
    of NGC~4151, grouped to require 20 counts per bin.  Error bars are
    omitted for visual clarity.  No models have been applied to the
    data, but the instrumental effecitive area function (ancillary
    response file, or ARF) has been removed. The spectra can
    be placed into two broad categories: those that have a higher flux
    and/or lower obscuration, and those that have a lower flux and/or
    higher obscuration.  Observation identification numbers (ObsIDs)
    0335 (black), 7829 (blue), 16089 (magenta), and 16090 (yellow)
    have a lower intrinsic flux and/or higher obscuration.  ObsIDs
    3480 (red), 3052 (green), and 7830 (cyan) have a higher intrinsic
    flux and/or less obscuration).}
\end{figure}
\medskip

\clearpage

\begin{figure}
\hspace{0.75in}
\includegraphics[scale=0.75]{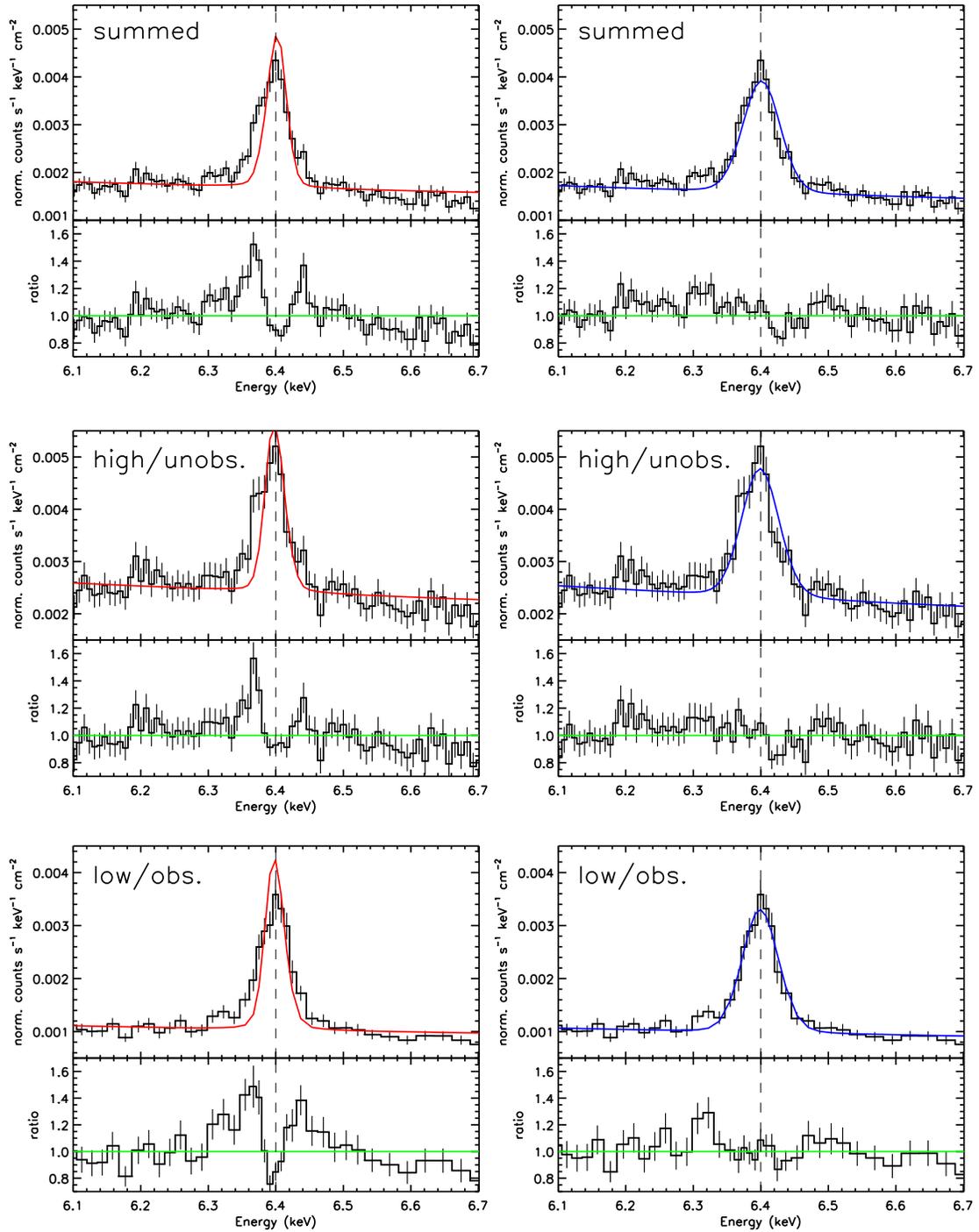}
\vspace{-0.5in}
  \figcaption[t]{\footnotesize First-order {\it Chandra}/HEG spectra
    of NGC~4151 in the Fe~K band.  The spectra have been shifted to
    the host frame.  The top, middle, and bottom rows show spectra
    from the summed spectrum, ``high/unobscured'' phase, and
    ``low/obscured'' phase, grouped to 20, 20, and 100 counts per bin,
    respectively.  The lefthand column shows fits with a simple
    Gaussian function with only instrumental broadening.  The
    righthand column shows fits with a Gaussian function with a
    variable width, $\sigma$.  In all cases, the energy of the
    Gaussian was fixed at $E=6.40$~keV, the laboratory energy for
    Fe~K$\alpha$.  These fits show that the line is clearly
    asymmetric, likely due to a combination of weak Doppler shifts and
    gravitational red-shifts, and scattering.  The results of these
    and other spectral fits are given in Tables 2 and 3.}
\end{figure}
\medskip

\begin{figure}
\hspace{0.75in}
\includegraphics[scale=0.75]{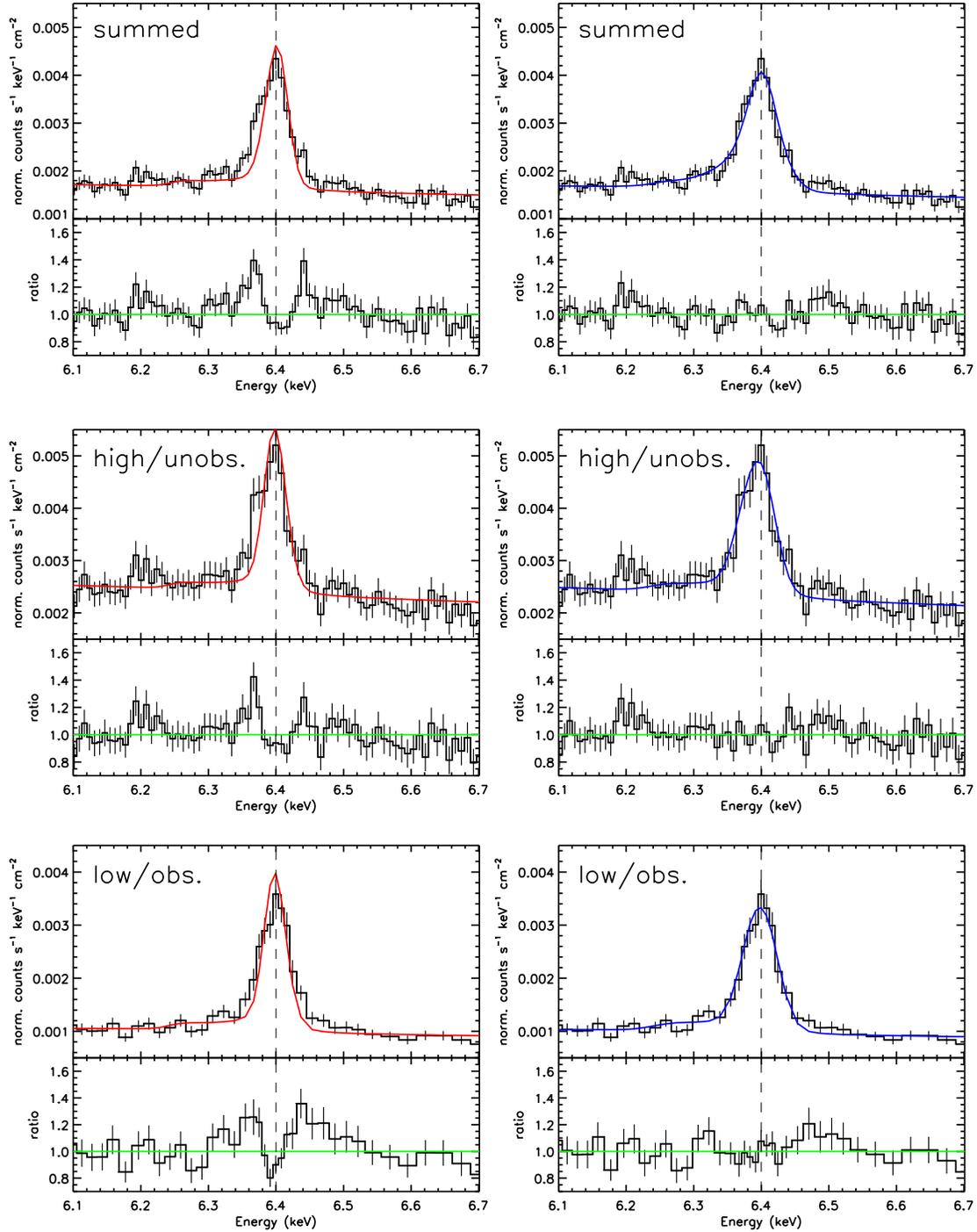}
\vspace{-0.5in}
  \figcaption[t]{\footnotesize Fits to the Fe~K$\alpha$ line with
    ``mytorus'' functions (see, e.g., Murphy \& Yaqoob 2009).  This
    function includes scattering effects in dense media that could
    potentially mimic dynamical broadening.  A subtle drop in the
    model at 6.25~keV, for instance, is the result of 150~eV energy loss from
    complete 180-degree Compton scattering.  Again, the top, middle,
    and bottom rows show spectra from the summed spectrum,
    ``high/unobscured'' phase, and ``low/obscured'' phase, grouped to
    20, 20, and 100 counts per bin, respectively.  The lefthand column
    shows simple fits with ``mytorus'' model.  The righthand column
    shows fits with the ``rdblur'' convolution function acting on
    ``mytorus'' to account for dynamics.  These fits are superior and
    demonstrate that dynamical broadening is likely required to
    describe the data.  The results of these and other spectral fits are
    given in Tables 2 and 3.}
\end{figure}
\medskip

\clearpage

\begin{figure}
\hspace{1.5in}
  \includegraphics[scale=0.6]{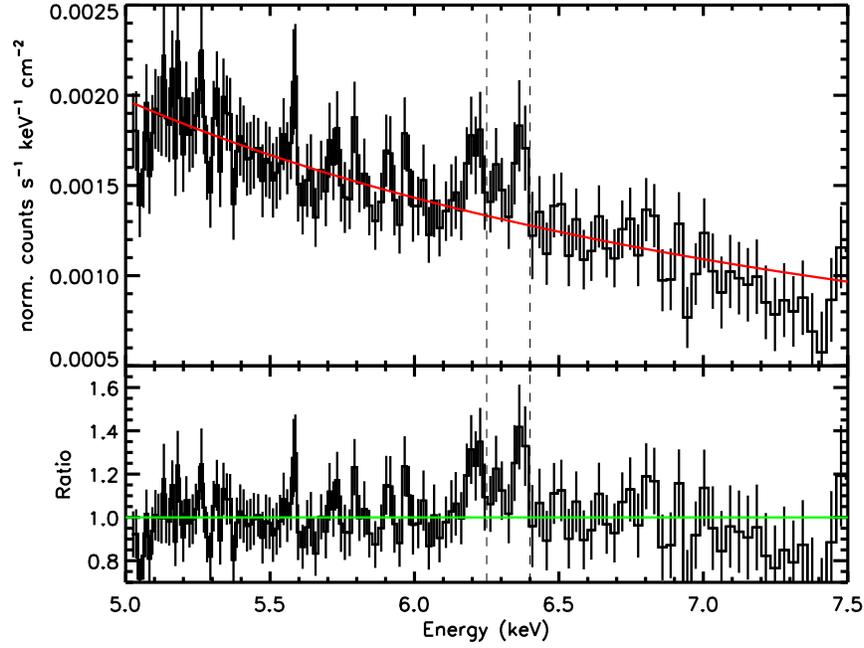}
  \figcaption[t]{\footnotesize The difference spectrum of NGC 4151,
    between the ``high/unobscured'' and ``low/obscured'' states.  The
    spectrum was created by subtracting the combined ``low/obscured''
    spectrum from the ``high/unobscured'' spectrum in units of count
    rate.  The remaining spectrum represents the variable flux between
    the two states.  The data were fit with a simple power-law model
    after grouping to require 20 counts per bin.  Dashed vertical
    lines at 6.40~keV and 6.25~keV mark the energy at which emission
    features are expected from neutral Fe~K$\alpha$, and the 150~eV
    energy loss incurred from 180-degree Compton scattering.  Emission
    lines are instead evident to the red of these energies, providing
    model-independent evidence that the red wing of the narrow
    Fe~K$\alpha$ line in NGC 4151 is stronger in the
    ``high/unobscured'' state.  When fit with simple Gaussian
    functions, the features at 6.37~keV and 6.20~keV are significant
    at the 4$\sigma$ and 5$\sigma$ level of confidence, respectively.
    Please see the text for additional details.}
\end{figure}
\medskip

\clearpage

\begin{figure}
\hspace{0.75in}
\includegraphics[scale=0.7]{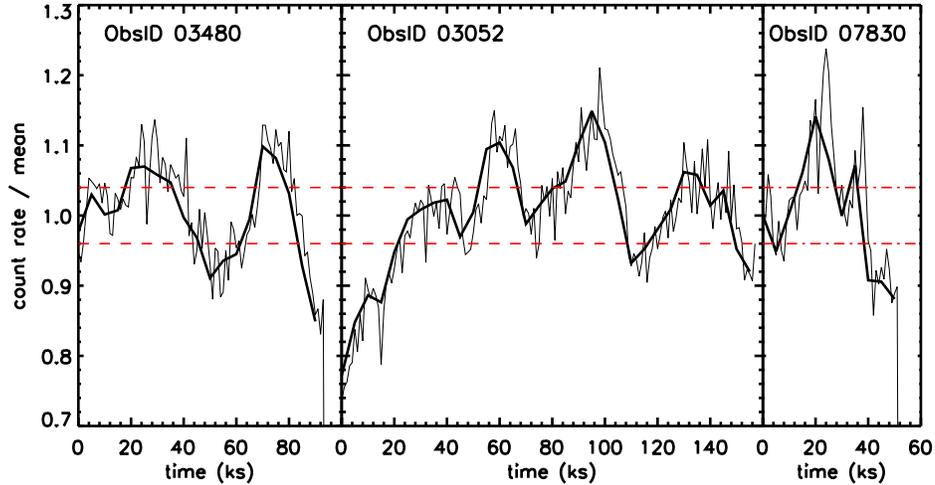}
\vspace{-0.75in}
  \figcaption[t]{\footnotesize The light curve of the observations
    obtained in the ``high/unobscured'' phase, wherein the
    Fe~K$\alpha$ emission line asymmetry is more pronounced and models
    measure smaller radii (see Tables 2 and 3).  The thick black curve
    has time bins of 5~ks; the thin gray line has time bins of 1~ks.
    The count rates are plotted as a ratio to the mean count rate in
    each observation (the mean rates are nearly identical between
    observations), to illustrate the relative magnitudes of the
    variations.  The red dashed lines mark flux levels $\pm4$\% above
    and below the mean in each observation.  To explore potential
    short time scale variations in the narrow Fe~K$\alpha$ line,
    spectra were extracted from intervals that differed by more than
    4\% from the mean.}
\end{figure}
\medskip

\begin{figure}
\hspace{1.0in}
  \includegraphics[scale=0.5,angle=-90]{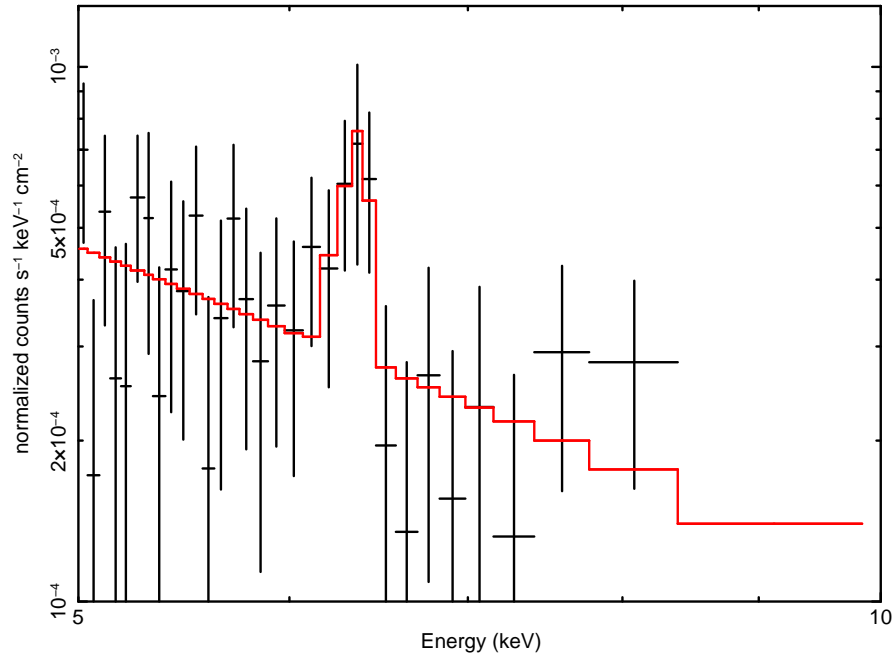}
  \figcaption[t]{\footnotesize The difference spectrum of NGC 4151 in
    the ``high/unobscured'' state, based on the count rate selections
    shown in Figure 5.  To create this spectrum, the spectrum of the
    low flux intervals was subtracted from the high flux intervals.
    Prior to fitting, the data were then grouped to require 50 counts
    per bin (further binning was done for visual clarity only).  The
    resulting difference spectrum is the variable flux spectrum.  The
    narrow Fe~K$\alpha$ line is weakly evident in the difference
    spectrum, suggesting that it may vary on the time scales typical
    of the flux intervals.  Here, the difference spectrum has been fit
    with a power-law continuum and a ``diskline'' model with an inner
    radius frozen at $R_{in} = 130~GM/c^{2}$.  Dividing the line flux
    by its minus-side error gives a significance of $3\sigma$.}
\end{figure}
\medskip

\clearpage

\begin{figure}
\hspace{2.0in}
\includegraphics[scale=0.48]{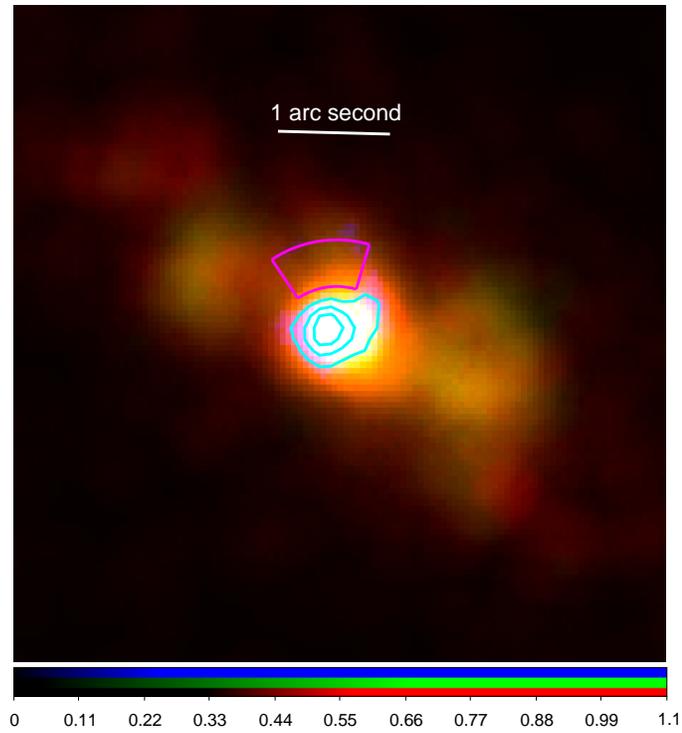}
\figcaption[t]{\footnotesize A ``true-color'' image of NGC 4151 from a
  {\it Chandra} imaging observation (ObsID 9217).  The observation was
  filtered to create sub-pixels of 0.0492 arc seconds, and further
  filtered to only include ``grade=0'' events in order to limit
  pile-up distortions (Miller et al.\ 2017).  Separate images were
  made in the 6.3-6.5 keV band (Fe~K, shown in blue), 0.8-1.1 keV band
  (Ne IX-X, shown in green), and 0.5-0.8 keV (O VII-VIII, shown in
  red).  A magenta region marks the location of the {\it Chandra} PSF
  artifact.  Cyan contours show that the Fe K region is largely
  contained within a radius of approximately 0.4 arc seconds, and is
  consistent with a point source.  In contrast, the soft X-ray
  emission from He-like and H-like charge states is extended and
  consistent with the optical NLR, as per Wang et al.\ (2011).  The
  narrow Fe~K$\alpha$ line is nuclear, and does not arise in the
  extended NLR.}
\end{figure}
\medskip

\clearpage

\begin{figure}
\hspace{0.05in}
\includegraphics[scale=0.48]{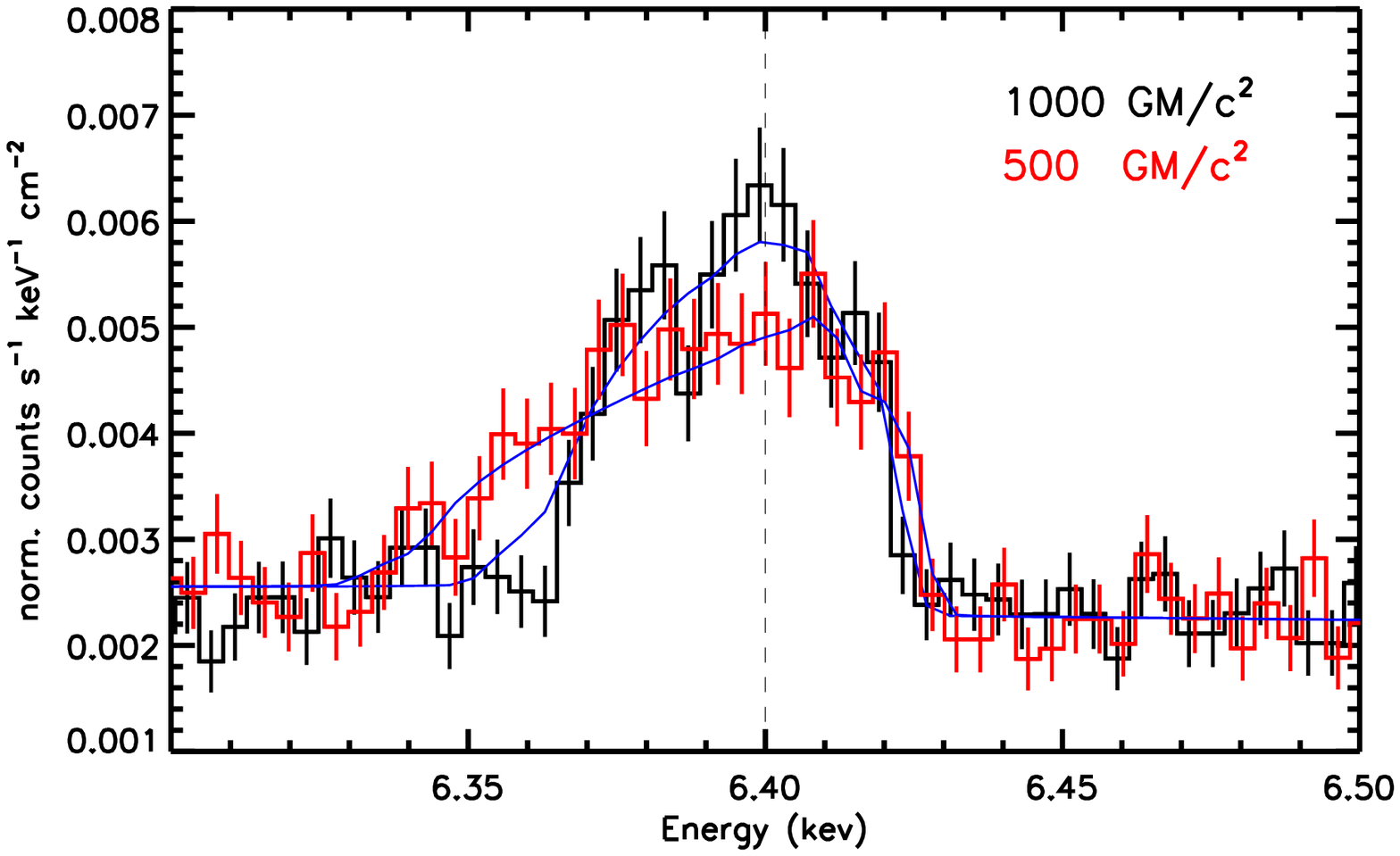}
  \hspace{-0.25in}
  \includegraphics[scale=0.48]{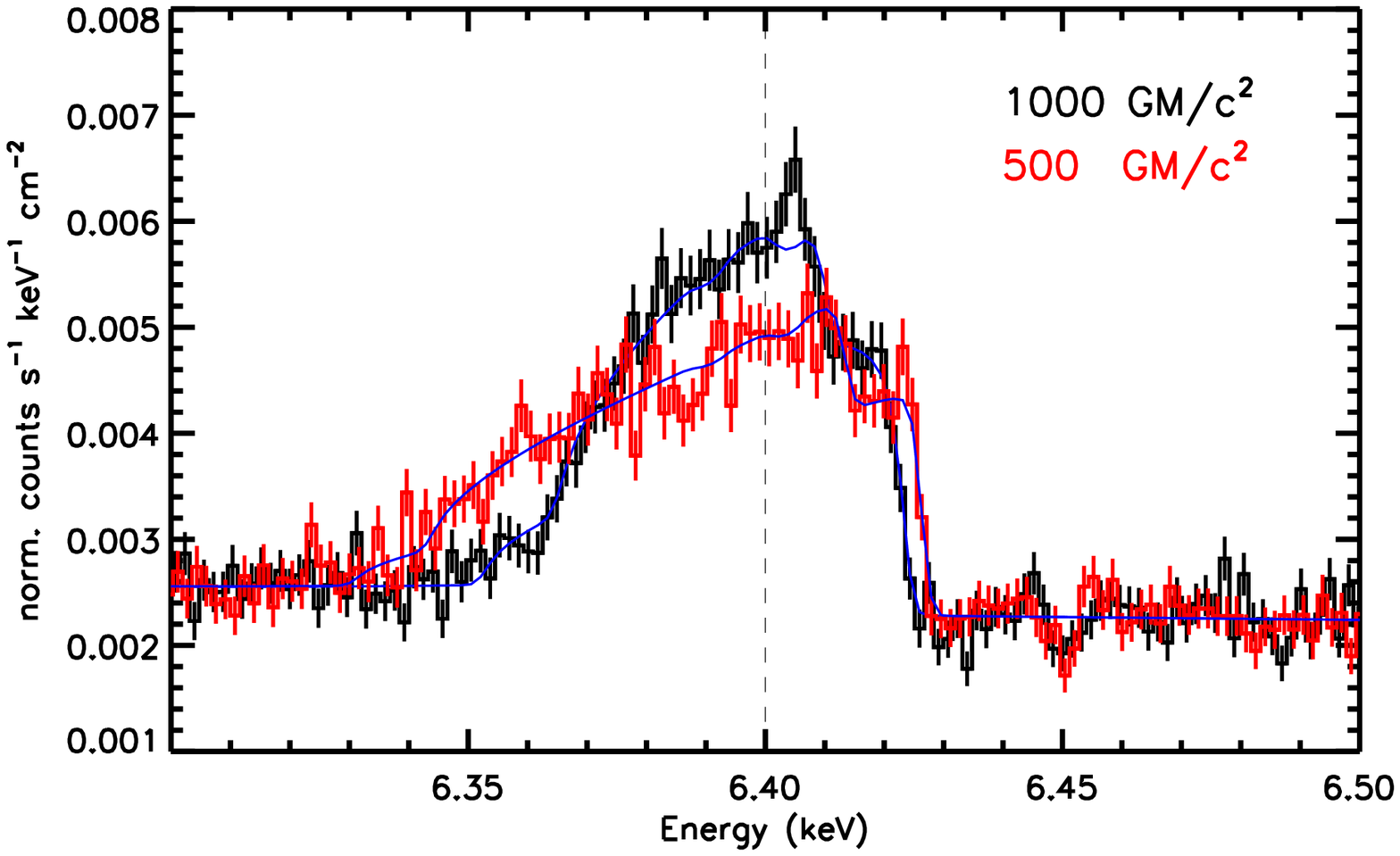}
  \vspace{-1.25in}
  \figcaption[t]{\footnotesize Simulated 20~ks {\it XARM} (left) and
    {\it ATHENA} (right) spectra of NGC 4151.  The spectra are binned
    for visal clarity and shifted to the frame of the host.  The
    simulations were based on the best-fit blurred ``mytorus'' models
    for in Table 2.  Emission originating from $R \geq 500~GM/c^{2}$
    (red) and $R\geq 1000~GM/c^{2}$ (black) was simulated for both
    missions.  The spectra illustrate the power of calorimeter
    resolution to easily distinguish even large radii through subtle
    changes to weak Doppler boosting in Keplerian orbits.  Calorimeter
    spectrometers may enable reverberation mapping from warps at
    intermediate disk radii and/or the BLR.}
\end{figure}
\medskip

\end{document}